\documentclass[pdflatex,sn-mathphys-num]{sn-jnl}

\usepackage{graphicx}
\usepackage{amsmath,amssymb,amsfonts}
\usepackage{mathtools}
\usepackage[title]{appendix}
\usepackage{placeins}
\usepackage{xurl}
\usepackage{enumitem}
\usepackage{booktabs}
\usepackage{tabularx}
\usepackage{xcolor}
\usepackage{tikz}
\usetikzlibrary{arrows.meta}
\usepackage{textcomp}
\usepackage{manyfoot}
\newcolumntype{Y}{>{\raggedright\arraybackslash}X}
\newcolumntype{L}[1]{>{\raggedright\arraybackslash}p{#1}}
\usepackage{etoolbox}
\hypersetup{hypertexnames=false, colorlinks=true, linkcolor=blue, citecolor=blue, urlcolor=blue, filecolor=blue, linktoc=all, pdftitle={Electromagnetic Radiation from a Neutralized Polarized Sphere with Two Conserved Currents for One Charge History}, pdfauthor={Natan Rentzber}, pdfkeywords={Classical Electrodynamics, Nonradiating Sources, Continuity Equation, Finite-Size Form Factor, Nonuniqueness}}

\makeatletter
\patchcmd{\@maketitle}{Corresponding author(s). E-mail(s): }{E-mail: }{}{\ClassError{sn-jnl-patch}{e-mail line patch failed}{}}
\def\email#1{\global\advance\emailcnt by 1\relax%
\if@corauemail
   \g@addto@macro\corrauthemail{\setcounter{footnote}{0}\textcolor{blue}{#1}}%
\else
   \g@addto@macro\authemail{\setcounter{footnote}{0}\textcolor{blue}{#1}}%
\fi}
\makeatother

\newcommand{\vE}{\mathbf{E}}
\newcommand{\vB}{\mathbf{B}}
\newcommand{\vD}{\mathbf{D}}
\newcommand{\vH}{\mathbf{H}}
\newcommand{\vP}{\mathbf{P}}
\newcommand{\vM}{\mathbf{M}}
\newcommand{\vJ}{\mathbf{J}}
\newcommand{\vK}{\mathbf{K}}
\newcommand{\vT}{\mathbf{T}}
\newcommand{\vA}{\mathbf{A}}
\newcommand{\vS}{\mathbf{S}}

\newcommand{\vx}{\mathbf{x}}
\newcommand{\vp}{\mathbf{p}}
\newcommand{\zero}{\mathbf{0}}
\newcommand{\nhat}{\hat{\mathbf{n}}}

\newcommand{\rhat}{\hat{\mathbf{r}}}
\newcommand{\zhat}{\hat{\mathbf{z}}}

\newcommand{\vq}{\mathbf{q}}

\newcommand{\shat}{\hat{\mathbf{s}}}

\newcommand{\thetahat}{\hat{\boldsymbol{\theta}}}
\newcommand{\phihat}{\hat{\boldsymbol{\phi}}}
\newcommand{\dd}{\,\mathrm d}
\newcommand{\divs}{\nabla_{\!s}\!\cdot}

\allowdisplaybreaks

\begin{document}

\title[Two Conserved Currents for One Charge History]{Electromagnetic Radiation from a Neutralized Polarized Sphere with Two Conserved Currents for One Charge History}

\author*[1]{\fnm{Natan} \sur{Rentzber}}\email{nrentzbe@uccs.edu}

\affil*[1]{\orgdiv{Center for Magnetism and Magnetic Nanostructures},

\orgname{University of Colorado},
\orgaddress{\city{Colorado Springs}, \postcode{80918}, \country{USA}}}

\abstract{Can a source radiate when its total charge density vanishes identically? Consider a uniformly polarized sphere coated with free surface charge that cancels the bound surface charge at every point and time. Then $\rho_{\mathrm{tot}}=0$, so every electric charge multipole vanishes. Continuity determines only $\nabla\cdot\vJ$, which allows the same charge history to be supported by different conserved currents. A compensating interior current gives $\vJ_{\mathrm{tot}}=\mathbf 0$ and produces no $\vE$ or $\vB$ at any frequency. The minimum-norm tangential sheet current instead leaves $\vJ_{\mathrm{tot}}$ nonzero and divergence-free. Its radiation-zone field is exact in $kR$ and proportional to $j_2(kR)$. At long wavelength the radiated power is suppressed by $(kR)^4/100$, and it vanishes exactly at the positive roots of $j_2$. The same calculation gives the interior field and a closed-form energy balance. The average work supplied by driving equals the radiated power and falls to zero at those roots even though interior fields remain. For comparison, the bare sphere has the factor $3j_1(kR)/(kR)$ and is silent at the roots of $j_1$.}


\maketitle

\clearpage
\setcounter{tocdepth}{2}
\tableofcontents
\clearpage

\section{Introduction}
\label{sec:intro}

A uniformly polarized sphere is coated with free surface charge that cancels its bound surface charge at every instant. Once the polarization begins to oscillate, the answer to whether it radiates depends on the current that transports the charge. Charge conservation fixes $\nabla\cdot\vJ$ but leaves the solenoidal part undetermined. Two conserved currents are constructed for the same prescribed charge history. One produces no $\vE$ or $\vB$, while the other radiates through a single closed-form channel.
\\

Nonradiating accelerated sources have a long history. In 1933, Schott showed that a uniformly charged spherical shell of radius $a$, moving rigidly with period $T$, emits no radiation when $2a=mcT$ for a positive integer $m$~\cite{schott}. Bohm and Weinstein, followed by Goedecke, developed related examples~\cite{bohmweinstein,goedecke}. Devaney and Wolf later gave the exact condition for a compact monochromatic current. It is nonradiating when its transverse Fourier transform vanishes on the shell $|\vq|=\omega/c$~\cite{devaneywolf}. The same freedom appears in inverse-source problems. Nonradiating sources lie in the null space of the map from a source to its exterior field, which prevents exterior measurements from determining a general source uniquely~\cite{bleisteincohen,devaneysherman}. Gbur reviews this literature~\cite{gbur}. Other classical treatments appear in Refs.~\cite{meyervernet,abbottgriffiths,loh}. The subject also remains active in nanophotonics, where anapole excitations suppress far-field radiation~\cite{miroshnichenko}.
\\

Mansuripur and Jakobsen obtained the exact fields of a uniformly polarized sphere without the neutralizing coating~\cite{mansuripur2020}. The source considered here adds that coating. The current required to maintain its charge is not unique, and the radiation depends on how the charge is transported. For the minimal tangential sheet current, the radiation-zone amplitude has the closed-form factor $\tfrac32 j_2(kR)$ relative to a point-dipole reference. At long wavelength this amplitude is smaller than ordinary dipole radiation by two powers of $kR$. The exact nonradiating frequencies also depend on the current realization. The bare sphere is silent at the positive roots of $j_1$. The minimal sheet realization is silent at the roots of $j_2$, while the volume-current realization is silent at every frequency. The form factors $3j_1(kR)/(kR)$ and $\tfrac32 j_2(kR)$ are derived below and compared in Table~\ref{tab:realizations} and Fig.~\ref{fig:formfactors}. The closed form $\tfrac32 j_2(kR)$ for the minimal sheet realization and its exact comparison with the bare-sphere factor $3j_1(kR)/(kR)$ appear not to have been reported previously.
\\

The static problem is considered first. Before retardation enters, the neutralized body has no force field outside or inside, while the auxiliary field $\vD$ remains confined to its interior. The dynamic boundary-value problem then gives the interior and exterior coefficients. For either radiating case, the exterior field coincides exactly with that of a point dipole whose moment carries the appropriate form factor. The energy balance is obtained in the same closed form. The average work supplied by the driving agent equals the radiated power and vanishes at the silent frequencies even though interior fields persist.
\newpage

Pointwise cancellation of $\rho_{\mathrm{tot}}$ keeps the charge history fixed while the two conserved current realizations are compared. The source is prescribed macroscopically and is not intended as a passive radiator. Section~\ref{sec:source} defines the two currents that support the same charge history and establishes the static baseline. Section~\ref{sec:fields} solves the field equations as a boundary-value problem. Section~\ref{sec:exact} obtains the exact interior and exterior fields, the radiated power, the energy balance, and the finite-size form factors. Section~\ref{sec:interpret} discusses the long-wavelength limit and the exact nonradiating frequencies.

\section{One Charge History, Two Conserved Currents}
\label{sec:source}

\subsection{Polarized Sphere and its Bound Sources}

Let a sphere of radius $R$ carry a prescribed spatially uniform polarization along the symmetry axis,
\begin{equation}
\vP(\vx,t)=P_{s}(t)\,\zhat,
\qquad
P_{s}(t)=P_{0}\cos\omega t,
\qquad r<R,
\label{eq:P}
\end{equation}
with $\vP=\mathbf 0$ outside. The constant $P_{0}$ has the SI units of polarization. The polarization is not treated as a constitutive response. No relation of the form $\vP=\varepsilon_{0}\chi_{e}\vE$ is assumed. Instead, it is an impressed macroscopic source driven in phase throughout the body~\cite{griffiths,herczynski}.
\\

A polarized medium carries bound charge. Integrating the potential of a polarized volume by parts gives the standard bound densities,
\begin{equation}
\rho_{b}=-\nabla\cdot\vP,
\qquad
\sigma_{b}=\vP\cdot\shat ,
\label{eq:bounddefs}
\end{equation}
where $\shat$ is the outward unit normal at the boundary. The derivative acting on $\vP$ produces the volume term, while the integration boundary produces the surface term. Because $\vP$ in Eq.~\eqref{eq:P} is uniform inside the sphere, its divergence vanishes there. The bound volume charge is therefore zero, and all bound charge lies on the surface,
\begin{equation}
\rho_{b}=0 \quad (r<R),
\qquad
\sigma_{b}(\theta,t)=P_{s}(t)\cos\theta ,
\label{eq:bound}
\end{equation}
where $\shat=\hat{\mathbf r}$ and $\zhat\cdot\hat{\mathbf r}=\cos\theta$.
\\

A time-dependent polarization also carries a bound current,
\begin{equation}
\vJ_{b}(\vx,t)=\frac{\partial\vP}{\partial t}=\frac{\dd P_{s}}{\dd t}\,\zhat
\qquad (r<R),
\label{eq:Jb}
\end{equation}
which is uniform throughout the interior and directed along the symmetry axis. Equations~\eqref{eq:bound} and~\eqref{eq:Jb} satisfy charge conservation. The surface charge in Eq.~\eqref{eq:bound} changes at the same rate that the interior current in Eq.~\eqref{eq:Jb} transports charge to the boundary.

\subsection{Neutralizing the Charge}

A free surface charge is placed on the sphere to cancel the bound layer at every point and time,
\begin{equation}
\sigma_{f}(\theta,t)=-\sigma_{b}(\theta,t)=-P_{s}(t)\cos\theta .
\label{eq:sigmaf}
\end{equation}
\\
The two coincident layers have equal magnitude and opposite sign. Their total charge density therefore vanishes identically as a distribution,
\begin{equation}
\rho_{\mathrm{tot}}(\vx,t)=\bigl[\sigma_{b}+\sigma_{f}\bigr]\delta(r-R)=0
\qquad\text{for all }\vx\text{ and all }t.
\label{eq:rhozero}
\end{equation}
Equation~\eqref{eq:rhozero} is stronger than a statement of zero net charge. The monopole is only the first cancellation. Every multipole moment formed from the total charge density,
\begin{equation}
q_{\ell m}(t)=\int Y^{*}_{\ell m}(\theta,\phi)\,r^{\ell}\,\rho_{\mathrm{tot}}(\vx,t)\dd^{3}x ,
\label{eq:multipoles}
\end{equation}
vanishes for every $\ell$ and $m$ at every time because the integrand is zero pointwise. In particular, the dipole moment is always zero. There is no electric dipole moment available to oscillate.
\\

The coincident sheets represent the zero-thickness limit of two matched layers with opposite signs. Cancellation at the level of distributions does not imply that the carriers annihilate. The bound and free layers remain physically distinct, although only their sum enters the Maxwell equations for $\vE$ and $\vB$.

\subsection{Current That Maintains the Neutral Surface}
\label{sec:continuity}

The surface charge in Eq.~\eqref{eq:sigmaf} changes with time and must be supplied by a current. Charge conservation on a patch of the interface gives the required balance. Charge accumulates at the rate $\partial\sigma_{f}/\partial t$. Tangential current leaves the patch at the rate $\divs\vK_{f}$, while volume current reaches it from the two sides. The surface-continuity equation is therefore
\\
\begin{equation}
\frac{\partial\sigma_{f}}{\partial t}+\divs\vK_{f}
=\bigl(\vJ_{f}^{\mathrm{in}}-\vJ_{f}^{\mathrm{out}}\bigr)\cdot\shat ,
\label{eq:surfacecontinuity}
\end{equation}
\\
where $\vK_{f}$ is the free surface-current density, measured as current per unit length. The quantities $\vJ_{f}^{\mathrm{in}}$ and $\vJ_{f}^{\mathrm{out}}$ are the one-sided limits of the free volume current, and $\nabla_{\!s}=(\mathbf I-\shat\shat)\cdot\nabla$ is the surface gradient.
\\

Equation~\eqref{eq:surfacecontinuity} supplies only one scalar condition and cannot determine the two components of $\vK_{f}$. For two currents $\vJ_{1}$ and $\vJ_{2}$ that support the same prescribed charge history, subtraction of their continuity equations gives
\begin{equation}
\nabla\cdot\bigl(\vJ_{2}-\vJ_{1}\bigr)=0 .
\label{eq:currentfreedom}
\end{equation}
Continuity determines the longitudinal part of the current but leaves any divergence-free addition unconstrained. The two realizations below make this freedom explicit.

\subsection{Realization A. Cancellation by an Interior Volume Current}

The first realization places all free current in the volume. It has $\vK_{f}=\mathbf 0$, no current outside, and
\\
\begin{equation}
\vJ_{f}^{\mathrm{in}}=-\frac{\dd P_{s}}{\dd t}\,\zhat
\qquad (r<R).
\label{eq:volumechoice}
\end{equation}
\\
Substituting Eq.~\eqref{eq:volumechoice} into Eq.~\eqref{eq:surfacecontinuity}, the normal component at the inner face of the surface is
\\
\begin{equation}
\vJ_{f}^{\mathrm{in}}\cdot\shat=-\frac{\dd P_{s}}{\dd t}\cos\theta
=\frac{\partial\sigma_{f}}{\partial t},
\label{eq:normalcheck}
\end{equation}
\\
which satisfies Eq.~\eqref{eq:surfacecontinuity} when $\vK_{f}=\mathbf 0$. This current is therefore admissible.
\\

In this realization, the cancellation is complete. The free volume current in Eq.~\eqref{eq:volumechoice} is the negative of the bound polarization current in Eq.~\eqref{eq:Jb}. The free surface charge in Eq.~\eqref{eq:sigmaf} is likewise the negative of the bound surface charge in Eq.~\eqref{eq:bound}. Both totals vanish,
\\
\begin{equation}
\rho_{\mathrm{tot}}=0
\qquad\text{and}\qquad
\vJ_{\mathrm{tot}}=\mathbf 0 ,
\label{eq:fourzero}
\end{equation}
\\
as distributions throughout space and for all time. Under outgoing retarded boundary conditions, with no incident wave and no independently added homogeneous field, identically vanishing charge and current densities produce no retarded $\vE$ or $\vB$ anywhere at any frequency.

\subsection{Realization B. Transport Along the Surface}
\label{sec:realB}

The second realization places no free current in the volume. The neutralizing charge is transported entirely along the coating. With $\vJ_{f}^{\mathrm{in}}=\vJ_{f}^{\mathrm{out}}=\mathbf 0$, Eq.~\eqref{eq:surfacecontinuity} becomes
\begin{equation}
\divs\vK_{f}=-\frac{\partial\sigma_{f}}{\partial t}
=\frac{\dd P_{s}}{\dd t}\cos\theta .
\label{eq:divKreq}
\end{equation}
One equation cannot determine a two-component tangential field. Decompose $\vK_{f}$ into a part that carries the required divergence and a divergence-free part,
\begin{equation}
\vK_{f}=\nabla_{\!s}\Psi+\vK_{\perp},
\qquad
\divs\vK_{\perp}=0 ,
\label{eq:surfaceHodge}
\end{equation}
where $\Psi$ is a scalar field on the sphere. Equation~\eqref{eq:surfaceHodge} follows from the completeness of $\nabla_{\!s}Y_{\ell m}$ and $\shat\times\nabla_{\!s}Y_{\ell m}$ for smooth tangential fields on a sphere when $\ell\geq1$. The same basis appears in Eq.~\eqref{eq:Kperpexpand} and Sec.~\ref{sec:fields}. A sphere has no harmonic tangential field, so a third term is unnecessary. The quantity $\Psi$ is a scalar potential for the surface current, not an electromagnetic potential, and it has units of current. The symbol $\Phi$ is therefore reserved for the electromagnetic scalar potential. For the outgoing Lorenz-gauge solution used here, that potential vanishes because $\rho_{\mathrm{tot}}=0$ everywhere, as shown in Sec.~\ref{sec:fields}. Only the gradient term contributes to Eq.~\eqref{eq:divKreq}. Taking the surface divergence of Eq.~\eqref{eq:surfaceHodge} and using $\divs\nabla_{\!s}=\nabla_{\!s}^{2}$ rewrites Eq.~\eqref{eq:divKreq} as a Poisson equation on the sphere,
\\
\begin{equation}
\nabla_{\!s}^{2}\Psi=\frac{\dd P_{s}}{\dd t}\cos\theta .
\label{eq:surfpoisson}
\end{equation}
\\
Equation~\eqref{eq:surfpoisson} can be expanded in the eigenfunctions of the surface Laplacian. On a sphere of radius $R$, the spherical harmonics satisfy
\begin{equation}
\nabla_{\!s}^{2}Y_{\ell m}=-\frac{\ell(\ell+1)}{R^{2}}\,Y_{\ell m} .
\label{eq:surflapeig}
\end{equation}
No assumption about the surviving modes is needed, so the scalar field is written as
\begin{equation}
\Psi(\theta,\phi,t)=\sum_{\ell\geq0}\sum_{m=-\ell}^{\ell}\Psi_{\ell m}(t)\,Y_{\ell m}(\theta,\phi),
\label{eq:Psiexpand}
\end{equation}
while the source on the right side of Eq.~\eqref{eq:surfpoisson} contains only one mode because $\cos\theta=\sqrt{4\pi/3}\,Y_{10}$,
\\
\begin{equation}
\frac{\dd P_{s}}{\dd t}\cos\theta=\sqrt{\frac{4\pi}{3}}\,\frac{\dd P_{s}}{\dd t}\,Y_{10}(\theta,\phi).
\label{eq:srcexpand}
\end{equation}
\\
Substitution of Eq.~\eqref{eq:Psiexpand} into Eq.~\eqref{eq:surfpoisson}, followed by termwise use of Eq.~\eqref{eq:surflapeig}, gives
\begin{equation}
-\sum_{\ell m}\frac{\ell(\ell+1)}{R^{2}}\,\Psi_{\ell m}\,Y_{\ell m}(\theta,\phi)
=\sqrt{\frac{4\pi}{3}}\,\frac{\dd P_{s}}{\dd t}\,Y_{10}(\theta,\phi).
\label{eq:modesum}
\end{equation}
Multiplication by $Y^{*}_{\ell'm'}$ followed by integration over the sphere projects onto a single mode. The required orthonormality relation is
\begin{equation}
\int Y^{*}_{\ell'm'}(\theta,\phi)\,Y_{\ell m}(\theta,\phi)\dd\Omega=\delta_{\ell\ell'}\delta_{mm'} .
\label{eq:Yortho}
\end{equation}
The projection uses the conjugate at the same angular coordinates as the expansion. Orthogonality removes every term in Eq.~\eqref{eq:modesum} except the selected one. The right side remains only when $\ell'=1$ and $m'=0$, which leaves one algebraic equation for each mode,
\\
\begin{equation}
-\frac{\ell(\ell+1)}{R^{2}}\,\Psi_{\ell m}
=\sqrt{\frac{4\pi}{3}}\,\frac{\dd P_{s}}{\dd t}\,\delta_{\ell 1}\delta_{m0} .
\label{eq:modebymode}
\end{equation}
\\
Equation~\eqref{eq:modebymode} determines all coefficients. For $\ell\geq2$, and for $\ell=1$ with $m\neq0$, the right side vanishes while the prefactor does not, so $\Psi_{\ell m}=0$. When $\ell=0$, both sides vanish and $\Psi_{00}$ remains arbitrary. This coefficient produces no current because $Y_{00}$ is constant and $\nabla_{\!s}Y_{00}=0$. It is the arbitrary additive constant of the scalar potential. 
\\

The only nonzero coefficient is
\begin{equation}
\Psi_{10}=-\frac{R^{2}}{2}\sqrt{\frac{4\pi}{3}}\,\frac{\dd P_{s}}{\dd t},
\qquad\text{so}\qquad
\Psi=\Psi_{10}Y_{10}=-\frac{R^{2}}{2}\frac{\dd P_{s}}{\dd t}\cos\theta .
\label{eq:Asolve}
\end{equation}
The prescribed charge history therefore selects a unique gradient-type sheet current, and it is a pure $\ell=1$, $m=0$ mode. For an axisymmetric field, $\nabla_{\!s}=R^{-1}\partial_{\theta}\,\thetahat$. Applying this surface gradient to Eq.~\eqref{eq:Asolve} gives
\begin{equation}
\vK_{f}(\theta,t)=\nabla_{\!s}\Psi
=\frac{R}{2}\frac{\dd P_{s}}{\dd t}\sin\theta\,\thetahat .
\label{eq:Kf}
\end{equation}
The surface divergence of Eq.~\eqref{eq:Kf} is
\begin{equation}
\divs\vK_{f}=\frac{1}{R\sin\theta}\frac{\partial}{\partial\theta}
\Bigl(\sin\theta\cdot\frac{R}{2}\frac{\dd P_{s}}{\dd t}\sin\theta\Bigr)
=\frac{\dd P_{s}}{\dd t}\cos\theta ,
\label{eq:divK}
\end{equation}
in agreement with Eq.~\eqref{eq:divKreq}.
\\

The two current contributions form a closed loop. When $\dd P_{s}/\dd t>0$, the polarization current in Eq.~\eqref{eq:Jb} carries charge through the interior toward the north pole. The sheet current in Eq.~\eqref{eq:Kf} returns it along the surface from north to south. Together they form closed meridional circuits.
\\

\emph{Remaining Freedom.} The charge history does not fix the field $\vK_{\perp}$ in Eq.~\eqref{eq:surfaceHodge}. By construction, it has no surface divergence and therefore does not appear in Eq.~\eqref{eq:divKreq}. In the same basis it is the general surface-divergence-free field,
\begin{equation}
\vK_{\perp}=\sum_{\ell\geq1}\sum_{m=-\ell}^{\ell}b_{\ell m}(t)\,\bigl(\shat\times\nabla_{\!s}Y_{\ell m}\bigr),
\label{eq:Kperpexpand}
\end{equation}
with arbitrary coefficients $b_{\ell m}$. Equation~\eqref{eq:Kperpexpand} expresses the freedom in Eq.~\eqref{eq:currentfreedom} for currents confined to the coating. Adding any such field leaves $\sigma_{f}$ unchanged and gives another admissible current. The choice in Eq.~\eqref{eq:Kf} sets $\vK_{\perp}=\mathbf 0$ and is used throughout. It also minimizes $\oint|\vK_{f}|^{2}\dd S$. The minimum occurs at $\vK_{\perp}=\mathbf 0$ because the two terms in Eq.~\eqref{eq:surfaceHodge} are orthogonal under the surface inner product. Minimum-norm source selection is common in inverse-source theory, where it separates a chosen source from its nonradiating part~\cite{marengoziolkowski,marengojmp}.
\\

\emph{Total Current.} Combining Eqs.~\eqref{eq:Jb} and~\eqref{eq:Kf}, with the surface current written as a volume density through a radial delta function, gives
\begin{equation}
\vJ_{\mathrm{tot}}(\vx,t)=\frac{\dd P_{s}}{\dd t}
\Bigl[\zhat\,\Theta(R-r)+\frac{R}{2}\sin\theta\,\thetahat\,\delta(r-R)\Bigr].
\label{eq:Jtot}
\end{equation}
Because $\rho_{\mathrm{tot}}=0$ at all times, Eq.~\eqref{eq:Jtot} must be divergence-free. The volume contribution is
\\
\begin{equation}
\nabla\cdot\Bigl[\frac{\dd P_{s}}{\dd t}\zhat\,\Theta(R-r)\Bigr]
=-\frac{\dd P_{s}}{\dd t}\cos\theta\,\delta(r-R),
\label{eq:conscheck}
\end{equation}
\\
where the derivative of the step function produces the surface term. The sheet contributes $(\divs\vK_{f})\delta(r-R)=+(\dd P_{s}/\dd t)\cos\theta\,\delta(r-R)$ by Eq.~\eqref{eq:divK}, and the two terms cancel exactly.

\subsection{What the Two Realizations Share and Where They Differ}

Realizations A and B support the same charge history. In both cases $\rho_{\mathrm{tot}}=0$ everywhere and at every instant, Eq.~\eqref{eq:surfacecontinuity} is satisfied, and every charge multipole vanishes. The only difference lies in the divergence-free part of the current, which continuity cannot determine according to Eq.~\eqref{eq:currentfreedom}. Realization A has $\vJ_{\mathrm{tot}}=\mathbf 0$ and radiates nothing. Realization B has a nonzero divergence-free total current. Its radiation is derived in Sec.~\ref{sec:fields} and the sections that follow.

\begin{figure}[!htbp]
\centering
\resizebox{0.98\textwidth}{!}{%
\begin{tikzpicture}[>={Stealth[length=2.1mm,width=1.7mm]},line cap=round]
\begin{scope}
  \draw[very thick] (0,0) circle (1.3);
  \foreach \x in {-0.7,0,0.7} {
    \draw[->,thick] (\x-0.07,-0.72) -- (\x-0.07,0.72);
    \draw[->,thick,dashed] (\x+0.07,0.72) -- (\x+0.07,-0.72);
  }
\end{scope}
\node[align=center] at (0,-1.95) {\small (a) Volume Realization\\[2pt]
  \small $\vJ_f=-\vJ_b$,\quad $\rho_{\mathrm{tot}}=0$,\quad $\vJ_{\mathrm{tot}}=\mathbf 0$};
\begin{scope}[shift={(6.2,0)}]
  \draw[very thick] (0,0) circle (1.3);
  \foreach \x in {-0.6,0,0.6} \draw[->,thick] (\x,-0.72) -- (\x,0.72);
  \node at (0,1.0) {\small $\vJ_b$};
  \draw[->,thick] (70:1.46) arc[start angle=70,end angle=-70,radius=1.46];
  \draw[->,thick] (110:1.46) arc[start angle=110,end angle=250,radius=1.46];
  \draw[thin] (1.70,0.80) -- (1.40,0.60);
  \node[anchor=west] at (1.70,0.86) {\small $\vK_f$};
\end{scope}
\node[align=center] at (6.2,-1.95) {\small (b) Minimal Sheet Realization\\[2pt]
  \small $\rho_{\mathrm{tot}}=0$,\ $\nabla\!\cdot\vJ_{\mathrm{tot}}=0$,\ $\vJ_{\mathrm{tot}}\neq\mathbf 0$};
\node[align=center] at (3.1,2.05) {\small $\sigma_b=P_s\cos\theta$ and
  $\sigma_f=-\sigma_b$ coincide on $r=R$ in both realizations};
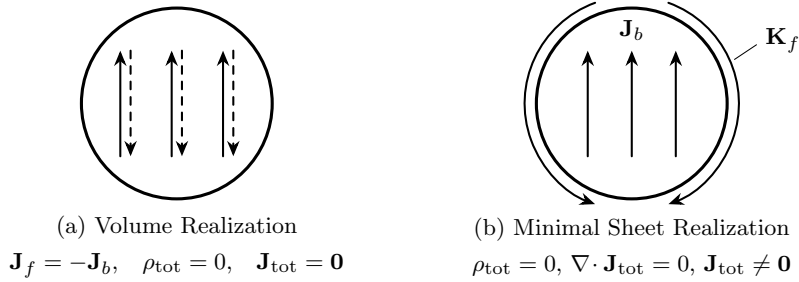
\node[align=center] at (3.1,-2.95) {\small current directions drawn for
  $\dd P_s/\dd t>0$, with paired arrows in (a) offset only for visibility};
\end{tikzpicture}%
}
\caption{Two currents supporting the same charge history. In both, the free sheet $\sigma_f$ cancels the bound sheet $\sigma_b$ at every point and time. (a)~In the volume-current realization, the compensating free current $\vJ_f=-\vJ_b$, shown by dashed arrows, cancels the polarization current $\vJ_b$, shown by solid arrows, at every interior point. The arrows are offset slightly so that both remain visible. (b)~In the minimal tangential-sheet realization, the polarization current fills the sphere and the meridional surface current $\vK_f$ from Eq.~\eqref{eq:Kf} carries the neutralizing charge from pole to pole along $\thetahat$. Its surface divergence satisfies Eq.~\eqref{eq:surfacecontinuity}, leaving a total current that is nonzero and divergence-free.}
\label{fig:source}
\end{figure}

\FloatBarrier
\subsection{Static Problem}
\label{sec:statics}

In the static limit, $\dd P_{s}/\dd t=0$, so both the polarization current in Eq.~\eqref{eq:Jb} and the sheet current of Sec.~\ref{sec:realB} vanish. The two realizations reduce to the same configuration. This case sets the $\omega\rightarrow0$ limit of the oscillating solution and shows the effect of neutralization before retardation appears.

The bare sphere carries only the bound surface charge in Eq.~\eqref{eq:bound}. Its electrostatic potential satisfies Poisson's equation $\nabla^{2}\Phi=-\rho_{\mathrm{tot}}/\varepsilon_{0}$~\cite{griffiths}. Since the bound volume charge in Eq.~\eqref{eq:bound} vanishes, Poisson's equation reduces to Laplace's equation on either side of the surface,
\\
\begin{equation}
\nabla^{2}\Phi=0 \qquad (r<R \ \text{ and } \ r>R),
\label{eq:laplace}
\end{equation}
\\
while the surface charge enters only through the matching conditions at $r=R$. The dynamic problem in Sec.~\ref{sec:fields} has the same boundary-driven structure.
\\

For the separated form $\Phi(r,\theta)=f(r)\,\Theta(\theta)$, substitution into Eq.~\eqref{eq:laplace} and multiplication by $r^{2}/(f\Theta)$ separate the radial and angular variables. Each group equals the constant $\ell(\ell+1)$. The angular equation is
\begin{equation}
\frac{1}{\sin\theta}\frac{\dd}{\dd\theta}\Bigl(\sin\theta\,\frac{\dd\Theta}{\dd\theta}\Bigr)+\ell(\ell+1)\,\Theta=0,
\label{eq:legendrezero}
\end{equation}
which is Legendre's equation with azimuthal index zero. Its solutions that remain regular on the polar axis are the Legendre polynomials $P_{\ell}(\cos\theta)$~\cite{dlmf}. The index is zero because the potential is a scalar. In Sec.~\ref{sec:fields}, the same separation applied to the azimuthal component of a vector field gives the associated Legendre equation of order one in Eq.~\eqref{eq:angular}. That order comes from the vector angular operator and does not imply azimuthal dependence.
\\

The radial equation is
\begin{equation}
r^{2}\frac{\dd^{2}f}{\dd r^{2}}+2r\frac{\dd f}{\dd r}-\ell(\ell+1)\,f=0,
\label{eq:euler}
\end{equation}
which is an equidimensional Euler equation. Every term carries the same power of $r$, so substitution of $f=r^{s}$ gives the indicial equation
\begin{equation}
s(s+1)=\ell(\ell+1),
\qquad
s=\ell \ \text{ or } \ s=-(\ell+1),
\label{eq:indicial}
\end{equation}
with two roots for each $\ell$. Regularity at the origin removes $r^{-(\ell+1)}$ from the interior solution. Decay at infinity removes $r^{\ell}$ from the exterior solution. The general axisymmetric potential is therefore
\begin{equation}
\Phi_{\mathrm{in}}=\sum_{\ell\geq0}A_{\ell}\,r^{\ell}P_{\ell}(\cos\theta),
\qquad
\Phi_{\mathrm{out}}=\sum_{\ell\geq0}\frac{B_{\ell}}{r^{\ell+1}}\,P_{\ell}(\cos\theta).
\label{eq:phigeneral}
\end{equation}

Two conditions join the interior and exterior solutions. The potential is continuous because the line integral of a finite $\vE$ across a layer of vanishing thickness goes to zero. A Gaussian pillbox of face area $\Delta A$ encloses the surface charge $\sigma\,\Delta A$. In the zero-height limit, only the flux through its two faces remains, and Gauss's law gives
\begin{equation}
\varepsilon_{0}\Bigl[\frac{\partial\Phi_{\mathrm{in}}}{\partial r}-\frac{\partial\Phi_{\mathrm{out}}}{\partial r}\Bigr]_{r=R}=\sigma(\theta).
\label{eq:phijump}
\end{equation}
Substitution of the expansions in Eq.~\eqref{eq:phigeneral} into both conditions, followed by multiplication by $P_{\ell'}(\cos\theta)\sin\theta$ and integration over $\theta$, uses the orthogonality relation $\int_{-1}^{1}P_{\ell}P_{\ell'}\dd u=2\,\delta_{\ell\ell'}/(2\ell+1)$~\cite{dlmf} and separates each value of $\ell$. For $\ell\neq1$, the surface charge $\sigma_{b}=P_{s}\cos\theta=P_{s}P_{1}(\cos\theta)$ has no projection. The resulting homogeneous system has determinant
\begin{equation}
\det\begin{pmatrix} R^{\ell} & -R^{-(\ell+1)}\\[2pt] \ell R^{\ell-1} & (\ell+1)R^{-(\ell+2)}\end{pmatrix}=\frac{2\ell+1}{R^{2}},
\label{eq:staticdet}
\end{equation}
which never vanishes. Hence $A_{\ell}=B_{\ell}=0$ for every $\ell\neq1$. The same mechanism removes the $\ell\neq1$ channels in the dynamic problem. The surviving equations are
\begin{equation}
A_{1}R=\frac{B_{1}}{R^{2}},
\qquad
\varepsilon_{0}\Bigl(A_{1}+\frac{2B_{1}}{R^{3}}\Bigr)=P_{s},
\label{eq:staticpair}
\end{equation}
and elimination of either coefficient gives
\begin{equation}
A_{1}=\frac{P_{s}}{3\varepsilon_{0}},
\qquad
B_{1}=\frac{P_{s}R^{3}}{3\varepsilon_{0}}.
\label{eq:staticcoeffs}
\end{equation}

The coefficients determine the fields. Inside, $\Phi_{\mathrm{in}}=(P_{s}/3\varepsilon_{0})\,r\cos\theta$ and $\nabla(r\cos\theta)=\zhat$, so
\begin{equation}
\vE_{\mathrm{in}}=-\nabla\Phi_{\mathrm{in}}=-\frac{P_{s}}{3\varepsilon_{0}}\,\zhat,
\label{eq:Einstatic}
\end{equation}
\\
which is uniform and antiparallel to the polarization. Outside, the potential is the field of an ideal point dipole carrying the full moment of the polarized body,
\begin{equation}
\vp_{b}=\int_{r<R}\vP\,\dd^{3}x=\frac{4\pi}{3}R^{3}P_{s}\,\zhat,
\qquad
\Phi_{\mathrm{out}}=\frac{P_{s}R^{3}}{3\varepsilon_{0}}\,\frac{\cos\theta}{r^{2}}=\frac{p_{b}\cos\theta}{4\pi\varepsilon_{0}r^{2}},
\label{eq:phiout}
\end{equation}
the standard result for a uniformly polarized sphere~\cite{griffiths}.
\\

The auxiliary field clarifies the boundary behavior. With $\vD=\varepsilon_{0}\vE+\vP$, Eqs.~\eqref{eq:P} and~\eqref{eq:Einstatic} give
\\
\begin{equation}
\vD_{\mathrm{in}}=\varepsilon_{0}\vE_{\mathrm{in}}+\vP=\frac{2}{3}P_{s}\,\zhat,
\qquad
\vD_{\mathrm{out}}=\varepsilon_{0}\vE_{\mathrm{out}},
\label{eq:Dstatic}
\end{equation}
\\
and the parallel and normal boundary components of both fields can now be written together. Evaluation of Eqs.~\eqref{eq:Einstatic} and~\eqref{eq:phiout} at $r=R$ gives
\begin{equation}
E_{\theta}^{\mathrm{out}}-E_{\theta}^{\mathrm{in}}=0,
\qquad
\varepsilon_{0}\bigl(E_{r}^{\mathrm{out}}-E_{r}^{\mathrm{in}}\bigr)=P_{s}\cos\theta=\sigma_{b},
\label{eq:Esurface}
\end{equation}
\begin{equation}
D_{r}^{\mathrm{out}}-D_{r}^{\mathrm{in}}=0,
\qquad
D_{\theta}^{\mathrm{out}}-D_{\theta}^{\mathrm{in}}=P_{s}\sin\theta=-\,\vP\cdot\thetahat\big|_{\mathrm{in}}.
\label{eq:Dsurface}
\end{equation}
\\
The tangential component of $\vE$ is continuous, while its normal component jumps by the bound surface charge. The normal component of $\vD$ is continuous because the bare sphere carries no free charge. Its tangential component inherits the discontinuity of $\vP$. Both values in the first part of Eq.~\eqref{eq:Dsurface} equal $\tfrac{2}{3}P_{s}\cos\theta$. The two fields therefore divide the parallel and normal boundary information in the expected way. With no current or magnetization, $\vB=\zero$ and $\vH=\vB/\mu_{0}=\zero$ everywhere.
\\

With the neutralizing coating included, The free charge in Eq.~\eqref{eq:sigmaf} cancels the bound layer pointwise, so Eq.~\eqref{eq:rhozero} holds. Poisson's equation is source free in both regions. With $\sigma_{\mathrm{tot}}=0$, Eq.~\eqref{eq:phijump} gives a continuous normal derivative as well. The only decaying solution of Laplace's equation that satisfies both continuity conditions is the trivial potential $\Phi$,
\begin{equation}
\Phi\equiv0,
\qquad
\vE=\vB=\vH=\zero \quad\text{everywhere}.
\label{eq:staticsilence}
\end{equation}
The auxiliary field remains nonzero inside the body,
\begin{align}
\vD&=\varepsilon_{0}\vE+\vP=\vP && (r<R),\notag\\
\vD&=\zero && (r>R),\notag\\
\nhat\cdot\bigl(\vD_{\mathrm{out}}-\vD_{\mathrm{in}}\bigr)&=-P_{s}\cos\theta=\sigma_{f},
\label{eq:Dneutral}
\end{align}
so the free coating sources $\vD$ even though all force fields vanish. The static neutralized sphere has no $\vE$, no $\vB$, and no $\vH$ anywhere. Its only field is $\vD$, which equals the impressed polarization inside and vanishes outside.
\\

The static configuration cannot distinguish the two current realizations because no current flows. Once the polarization oscillates, Eq.~\eqref{eq:sigmaf} requires the coating charge to change. That charge must be transported according to Eq.~\eqref{eq:surfacecontinuity}, and Sec.~\ref{sec:continuity} showed that the transporting current is not unique. The remaining sections determine how this choice changes the fields. In the limit $\omega\rightarrow0$, the dynamic solutions recover the static results above. The neutralized interior fields vanish as $(kR)^{2}$, while the bare interior field approaches $-P_{s}\,\zhat/(3\varepsilon_{0})$ exactly.

\section{Solution of the Field Equations}
\label{sec:fields}

\subsection{Equation to be Solved}

The harmonic convention $e^{-i\omega t}$ is used, with $k=\omega/c$. From Sec.~\ref{sec:source}, the phasor form of the total current in Eq.~\eqref{eq:Jtot} follows from $\dd P_{s}/\dd t\rightarrow-i\omega P_{0}$, while the total charge remains identically zero. Maxwell's two curl equations become
\begin{equation}
\nabla\times\vB=\mu_{0}\vJ_\omega-\frac{i\omega}{c^{2}}\vE ,
\qquad
\nabla\times\vE=i\omega\vB .
\label{eq:maxwellphasor}
\end{equation}
Taking the curl of the first equation and substituting the second gives a single equation for $\vB$,
\\
\begin{equation}
\nabla\times\bigl(\nabla\times\vB\bigr)
=\mu_{0}\nabla\times\vJ_\omega-\frac{i\omega}{c^{2}}\bigl(i\omega\vB\bigr)
=\mu_{0}\nabla\times\vJ_\omega+k^{2}\vB .
\label{eq:curlcurl}
\end{equation}
\\
The identity $\nabla\times(\nabla\times\vB)=\nabla(\nabla\cdot\vB)-\nabla^{2}\vB$, together with $\nabla\cdot\vB=0$, reduces Eq.~\eqref{eq:curlcurl} to the driven vector Helmholtz equation
\begin{equation}
\bigl(\nabla^{2}+k^{2}\bigr)\vB=-\mu_{0}\,\nabla\times\vJ_\omega .
\label{eq:vhelm}
\end{equation}

The source term vanishes away from the interface. Inside the sphere, the free volume current is zero and the total volume current is the uniform bound vector $-i\omega P_{0}\zhat$, whose curl vanishes. No current exists outside. Thus
\begin{equation}
\nabla\times\vJ_\omega=\zero
\qquad\text{for } r<R \text{ and for } r>R ,
\label{eq:curlJzero}
\end{equation}
and Eq.~\eqref{eq:vhelm} is homogeneous in both bulk regions. The drive enters only at $r=R$, where derivatives of the step function and delta shell contribute through the matching conditions. This leaves a boundary-value problem driven at the interface.

\subsection{Field is Purely Azimuthal}

The symmetry of the source first determines which components of $\vB$ can appear. The potential formulation gives the result directly. In the Lorenz gauge, the phasor potentials satisfy~\cite{jackson,zangwill}
\\
\begin{equation}
\nabla\cdot\vA=\frac{i\omega}{c^{2}}\,\Phi,
\label{eq:lorenz}
\end{equation}
\\
and the gauge condition separates their wave equations,
\begin{equation}
\bigl(\nabla^{2}+k^{2}\bigr)\Phi=-\frac{\rho_{\mathrm{tot}}}{\varepsilon_{0}},
\qquad
\bigl(\nabla^{2}+k^{2}\bigr)\vA=-\mu_{0}\vJ_{\omega}.
\label{eq:waveeqs}
\end{equation}
Both retarded solutions use the same outgoing Green function. The scalar potential is
\begin{equation}
\Phi(\vx)=\frac{1}{4\pi\varepsilon_{0}}\int\frac{e^{ik|\vx-\vx'|}}{|\vx-\vx'|}\,\rho_{\mathrm{tot}}(\vx')\,\dd^{3}x',
\label{eq:phiret}
\end{equation}
and its integrand vanishes pointwise by Eq.~\eqref{eq:rhozero} at every frequency. Hence
\begin{equation}
\Phi\equiv0,
\qquad
\vE=i\omega\vA,
\qquad
\vB=\nabla\times\vA,
\label{eq:phizero}
\end{equation}
as anticipated in Sec.~\ref{sec:realB}. The vector potential alone carries the electromagnetic field and is given by
\\
\begin{equation}
\vA(\vx)=\frac{\mu_0}{4\pi}\int \frac{e^{ik|\vx-\vx'|}}{|\vx-\vx'|}\,\vJ_\omega(\vx')\,\dd^3x',
\label{eq:Aret}
\end{equation}
where the integral acts component by component and preserves the symmetry of the source~\cite{jackson,zangwill}. The current in Sec.~\ref{sec:source} is axisymmetric and has only $\rhat$ and $\thetahat$ components. The vector potential $\vA_\omega$ is therefore axisymmetric with $A_{\phi}=0$ and $\partial_{\phi}=0$. Under these conditions, two components of its curl vanish identically,
\begin{equation}
(\nabla\times\vA)_{r}=\frac{1}{r\sin\theta}\Bigl[\partial_{\theta}\bigl(\sin\theta\,A_{\phi}\bigr)-\partial_{\phi}A_{\theta}\Bigr]=0,
\qquad
(\nabla\times\vA)_{\theta}=\frac{1}{r}\Bigl[\frac{\partial_{\phi}A_{r}}{\sin\theta}-\partial_{r}\bigl(rA_{\phi}\bigr)\Bigr]=0 ,
\label{eq:curlcomp}
\end{equation}
because every term contains either $A_{\phi}$ or a derivative with respect to $\phi$. Only the azimuthal component remains,
\begin{equation}
\vB=B_{\phi}(r,\theta)\,\phihat .
\label{eq:Bazim}
\end{equation}
The magnetic field is therefore determined by one scalar function of the radial and polar coordinates.

\subsection{Separation of Variables}

For the field in Eq.~\eqref{eq:Bazim}, the divergence vanishes identically and the vector Laplacian becomes $\nabla^{2}\vB=-\nabla\times(\nabla\times\vB)$. Evaluating the two curls and selecting the azimuthal component gives, in either source-free region where Eq.~\eqref{eq:vhelm} is homogeneous,
\begin{equation}
\frac{1}{r^{2}}\frac{\partial}{\partial r}\Bigl(r^{2}\frac{\partial B_{\phi}}{\partial r}\Bigr)
+\frac{1}{r^{2}\sin\theta}\frac{\partial}{\partial\theta}\Bigl(\sin\theta\frac{\partial B_{\phi}}{\partial\theta}\Bigr)
-\frac{B_{\phi}}{r^{2}\sin^{2}\theta}
+k^{2}B_{\phi}=0 .
\label{eq:helmexp}
\end{equation}
The final term distinguishes Eq.~\eqref{eq:helmexp} from the scalar Helmholtz equation. It appears because the vector Laplacian acting on a $\phihat$-directed field differentiates the direction of $\phihat$ as well as the magnitude of $B_{\phi}$. This term fixes the order of the angular functions.
\\

A separated form is
\\
\begin{equation}
B_{\phi}(r,\theta)=b(r)\,\Theta(\theta)
\label{eq:sepform}
\end{equation}
\\
Substitution into Eq.~\eqref{eq:helmexp}, followed by multiplication by $r^{2}/\bigl(b\Theta\bigr)$, places all $r$ dependence in one group and all $\theta$ dependence in the other,
\begin{equation}
\underbrace{\frac{1}{b}\frac{\dd}{\dd r}\Bigl(r^{2}\frac{\dd b}{\dd r}\Bigr)+k^{2}r^{2}}_{\text{depends only on }r}
\;+\;
\underbrace{\frac{1}{\Theta}\Bigl[\frac{1}{\sin\theta}\frac{\dd}{\dd\theta}\Bigl(\sin\theta\frac{\dd\Theta}{\dd\theta}\Bigr)-\frac{\Theta}{\sin^{2}\theta}\Bigr]}_{\text{depends only on }\theta}
\;=\;0 .
\label{eq:separated}
\end{equation}
The two groups depend on independent variables and sum to zero, so each equals a constant. Write the separation constant as $\ell(\ell+1)$ with the conventional sign that gives the standard angular solutions.

\subsection{Angular Equation}

The angular part of Eq.~\eqref{eq:separated} is
\begin{equation}
\frac{1}{\sin\theta}\frac{\dd}{\dd\theta}\Bigl(\sin\theta\frac{\dd\Theta}{\dd\theta}\Bigr)
+\Bigl[\ell(\ell+1)-\frac{1}{\sin^{2}\theta}\Bigr]\Theta=0 .
\label{eq:angular}
\end{equation}
This is the associated Legendre equation of order one. In the general equation, the corresponding term is $m^{2}/\sin^{2}\theta$. Here the $1/\sin^{2}\theta$ term below Eq.~\eqref{eq:helmexp} fixes the order at unity. The superscript does not imply explicit azimuthal phase dependence. Both the source and the field remain axisymmetric with $\partial_{\phi}=0$, so the spherical-harmonic azimuthal index remains zero. The order-one function arises from the vector angular operator acting on a field directed along $\phihat$.
\\

The solutions regular on the polar axis are
\begin{equation}
\Theta(\theta)=P^{1}_{\ell}(\cos\theta),
\qquad \ell=1,2,3,\dots,
\qquad
P^{1}_{1}(\cos\theta)=-\sin\theta ,
\label{eq:legendre}
\end{equation}
where $\ell$ begins at one because $P^{1}_{0}$ vanishes identically. These functions obey the orthogonality relation
\begin{equation}
\int_{0}^{\pi}P^{1}_{\ell}(\cos\theta)\,P^{1}_{\ell'}(\cos\theta)\,\sin\theta\dd\theta
=\frac{2\,\ell(\ell+1)}{2\ell+1}\,\delta_{\ell\ell'} ,
\label{eq:orthogonality}
\end{equation}
and Eq.~\eqref{eq:orthogonality} later separates the boundary conditions by multipole order~\cite{dlmf}.

\subsection{Radial Equation}

The radial part of Eq.~\eqref{eq:separated} is
\begin{equation}
\frac{\dd}{\dd r}\Bigl(r^{2}\frac{\dd b}{\dd r}\Bigr)+\bigl[k^{2}r^{2}-\ell(\ell+1)\bigr]b=0 ,
\label{eq:radial}
\end{equation}
or, after expansion and the substitution $x=kr$,
\begin{equation}
x^{2}\frac{\dd^{2}b}{\dd x^{2}}+2x\frac{\dd b}{\dd x}+\bigl[x^{2}-\ell(\ell+1)\bigr]b=0 .
\label{eq:sphbessel}
\end{equation}
Equation~\eqref{eq:sphbessel} is the spherical Bessel equation of order $\ell$. Its independent solutions are the spherical Bessel functions of the first and second kinds,
\begin{equation}
j_{\ell}(x),
\qquad
y_{\ell}(x),
\qquad
j_{1}(x)=\frac{\sin x}{x^{2}}-\frac{\cos x}{x},
\qquad
y_{1}(x)=-\frac{\cos x}{x^{2}}-\frac{\sin x}{x} ,
\label{eq:jydefs}
\end{equation}
where the order-one forms are displayed because only that order survives the boundary conditions. The recurrence and derivative identities used below are standard~\cite{dlmf}.

Regularity at the origin and outgoing behavior at infinity select the radial functions. Near the origin,
\begin{equation}
j_{\ell}(x)\longrightarrow\frac{x^{\ell}}{(2\ell+1)!!},
\qquad
y_{\ell}(x)\longrightarrow-\frac{(2\ell-1)!!}{x^{\ell+1}}
\qquad (x\rightarrow0),
\label{eq:smallx}
\end{equation}
so $y_{\ell}$ diverges and cannot appear inside a source-free neighborhood of the origin. The interior solution therefore contains only $j_{\ell}$. At large distance, the useful combinations are the spherical Hankel functions $h^{(1)}_{\ell}=j_{\ell}+iy_{\ell}$ and $h^{(2)}_{\ell}=j_{\ell}-iy_{\ell}$, with asymptotic forms
\begin{equation}
h^{(1)}_{\ell}(x)\longrightarrow(-i)^{\ell+1}\frac{e^{ix}}{x},
\qquad
h^{(2)}_{\ell}(x)\longrightarrow i^{\ell+1}\frac{e^{-ix}}{x}
\qquad (x\rightarrow\infty).
\label{eq:hankelasym}
\end{equation}
After restoring the time factor $e^{-i\omega t}$, $h^{(1)}_{\ell}$ carries $e^{i(kr-\omega t)}$ and represents an outgoing wave. The function $h^{(2)}_{\ell}$ represents an incoming wave. Since no radiation arrives from infinity, only $h^{(1)}_{\ell}$ is retained outside.

\subsection{General Solution}

Only the symmetry of the source has been used so far. Combining the angular functions in Eq.~\eqref{eq:legendre} with the radial solutions selected above gives the most general solution of Eq.~\eqref{eq:vhelm} that is regular at the origin and outgoing at infinity,
\begin{equation}
B_{\phi}^{\mathrm{in}}(r,\theta)=\sum_{\ell\geq1}A_{\ell}\,j_{\ell}(kr)\,P^{1}_{\ell}(\cos\theta),
\qquad
B_{\phi}^{\mathrm{out}}(r,\theta)=\sum_{\ell\geq1}C_{\ell}\,h^{(1)}_{\ell}(kr)\,P^{1}_{\ell}(\cos\theta).
\label{eq:generalsol}
\end{equation}
Each value of $\ell$ carries two unknown coefficients. The boundary conditions determine them independently. For every $\ell\neq1$, the resulting system is homogeneous with a nonzero determinant, so those coefficients vanish without being excluded in advance.

\subsection{Boundary Conditions}

Applying the interface conditions at $r=R$ also requires the electric field. Rearranging the first equation in Eq.~\eqref{eq:maxwellphasor} gives
\begin{equation}
\vE=-\frac{ic^{2}}{\omega}\bigl(\mu_{0}\vJ_\omega-\nabla\times\vB\bigr),
\label{eq:Efromcurl}
\end{equation}
and for the azimuthal magnetic field in Eq.~\eqref{eq:Bazim}, the relevant components of the curl are
\\
\begin{equation}
(\nabla\times\vB)_{r}=\frac{1}{r\sin\theta}\frac{\partial}{\partial\theta}\bigl(\sin\theta\,B_{\phi}\bigr),
\qquad
(\nabla\times\vB)_{\theta}=-\frac{1}{r}\frac{\partial}{\partial r}\bigl(rB_{\phi}\bigr).
\label{eq:curlB}
\end{equation}
\\
Outside the sphere, $\vJ_\omega=\zero$. Immediately inside, the current components are $J_{r}=-i\omega P_{0}\cos\theta$ and $J_{\theta}=+i\omega P_{0}\sin\theta$, obtained from $\zhat=\cos\theta\,\rhat-\sin\theta\,\thetahat$.
\\

The four interface conditions follow directly from Maxwell's equations~\cite{jackson}. A Gaussian pillbox of face area $\Delta A$ and vanishing height gives the normal conditions. Its side-wall flux disappears in the limit. Applied to $\nabla\cdot\vB=0$, it yields $\nhat\cdot(\vB_{\mathrm{out}}-\vB_{\mathrm{in}})\,\Delta A=0$. Applied to Gauss's law, it encloses the surface charge $\sigma_{\mathrm{tot}}\,\Delta A$. A rectangular loop of length $\Delta\ell$ in a plane containing $\nhat$ gives the tangential conditions. As the loop height vanishes, Faraday's law retains only the two long sides. The Amp\`ere-Maxwell law also encloses the sheet current $K\,\Delta\ell$. The resulting conditions are
\\
\begin{equation}
\nhat \cdot\bigl(\vB_{\mathrm{out}} - \vB_{\mathrm{in}}\bigr)=0,
\qquad
\nhat \times\bigl(\vB_{\mathrm{out}} - \vB_{\mathrm{in}}\bigr)=\mu_0\vK,
\label{eq:bcB}
\end{equation}
\\
\begin{equation}
\nhat \times\bigl(\vE_{\mathrm{out}} - \vE_{\mathrm{in}}\bigr)=\zero,
\qquad
\varepsilon_{0}\,\nhat \cdot\bigl(\vE_{\mathrm{out}} - \vE_{\mathrm{in}}\bigr)=\sigma_{\mathrm{tot}},
\label{eq:bcE}
\end{equation}
\\
where $\nhat=\rhat$ points outward.
\\

The same pillbox and loop give the corresponding conditions for the auxiliary fields. Since $\vM=\zero$, $\vD=\varepsilon_{0}\vE+\vP$ and $\vH=\vB/\mu_{0}$. The equation $\nabla\cdot\vD=\rho_{f}$ retains only the free surface charge. The equation $\nabla\times\vH=\vJ_{f}+\partial\vD/\partial t$ retains only the free sheet current because the bound sheet current $\vM\times\nhat$ vanishes with $\vM$. Thus
\begin{equation}
\nhat\cdot\bigl(\vD_{\mathrm{out}}-\vD_{\mathrm{in}}\bigr)=\sigma_{f},
\qquad
\nhat\times\bigl(\vH_{\mathrm{out}}-\vH_{\mathrm{in}}\bigr)=\vK_{f}.
\label{eq:bcDH}
\end{equation}
These expressions are not additional conditions. On the inner side, the normal component of $\vD$ differs from $\varepsilon_{0}$ times the normal component of $\vE$ by $\vP\cdot\nhat$. Substituting $\sigma_{b}=\vP\cdot\shat$ from Eq.~\eqref{eq:bounddefs} into the normal condition in Eq.~\eqref{eq:bcE}, together with $\sigma_{\mathrm{tot}}=\sigma_{b}+\sigma_{f}$, reproduces the first relation. Dividing the tangential condition in Eq.~\eqref{eq:bcB} by $\mu_{0}$ reproduces the second because the free sheet is the only sheet current. The two descriptions encode the same interface, with the bound sources absorbed into $\vD$ as in Sec.~\ref{sec:statics}. Only two independent conditions remain.
\\

\emph{Normal $\vB$:} Equation~\eqref{eq:Bazim} has no radial component, so this condition is satisfied identically and supplies nothing.
\\

\emph{Tangential $\vB$:} With $\rhat\times\phihat=-\thetahat$ and the sheet current of Eq.~\eqref{eq:Kf}, whose angular factor is $\sin\theta=-P^{1}_{1}(\cos\theta)$,
\begin{equation}
\sum_{\ell\geq1}\Bigl[C_{\ell}h^{(1)}_{\ell}(kR)-A_{\ell}j_{\ell}(kR)\Bigr]P^{1}_{\ell}(\cos\theta)
=-\frac{i\mu_{0}\omega P_{0}R}{2}\,P^{1}_{1}(\cos\theta).
\label{eq:bcBtan}
\end{equation}

\emph{Normal $\vE$:} The total surface charge vanishes by construction, so this condition reads $\nhat\cdot(\vE_{\mathrm{out}}-\vE_{\mathrm{in}})=0$. Evaluating it with Eqs.~\eqref{eq:Efromcurl} and~\eqref{eq:curlB}, while retaining the interior $J_{r}$, reproduces Eq.~\eqref{eq:bcBtan} exactly. The two conditions are not independent. Their equivalence reflects charge conservation. Neutrality of the total surface charge and the surface-continuity equation that supplies the sheet current express the same interface balance from the electric and magnetic sides.
\\

\emph{Tangential $\vE$:} This is the second independent condition. Writing $u_{\ell}(r)\equiv r\,b_{\ell}(r)$ for each radial function and retaining the interior $J_{\theta}$,
\begin{equation}
\sum_{\ell\geq1}\Bigl[u^{\mathrm{out}\prime}_{\ell}(R)-u^{\mathrm{in}\prime}_{\ell}(R)\Bigr]P^{1}_{\ell}(\cos\theta)
=-i\mu_{0}\omega P_{0}R\,P^{1}_{1}(\cos\theta).
\label{eq:bcEtan}
\end{equation}

\subsection{Orthogonality Separates the Multipoles}

The right sides of Eqs.~\eqref{eq:bcBtan} and~\eqref{eq:bcEtan} contain only $P^{1}_{1}$. Both the sheet current and the interior current derived in Sec.~\ref{sec:source} carry the angular factor $\sin\theta$, and $\sin\theta$ is a single associated Legendre function.
\\

Multiplication of Eqs.~\eqref{eq:bcBtan} and~\eqref{eq:bcEtan} by $P^{1}_{\ell'}(\cos\theta)\sin\theta$, followed by integration over $\theta$, removes every term except the one with matching order through Eq.~\eqref{eq:orthogonality}. With $C\equiv-i\mu_{0}\omega P_{0}$, define
\begin{equation}
f_{\ell}(r)=r\,j_{\ell}(kr),
\qquad
g_{\ell}(r)=r\,h^{(1)}_{\ell}(kr),
\label{eq:fgdef}
\end{equation}
which gives one uncoupled pair of equations for each $\ell$,
\begin{equation}
\begin{pmatrix} g_{\ell}(R) & -f_{\ell}(R)\\[2pt] g'_{\ell}(R) & -f'_{\ell}(R)\end{pmatrix}
\begin{pmatrix} C_{\ell}\\[2pt] A_{\ell}\end{pmatrix}
=\begin{pmatrix} CR^{2}/2\\[2pt] CR\end{pmatrix}\delta_{\ell 1} .
\label{eq:system}
\end{equation}
Each multipole order is independent. The system is inhomogeneous for $\ell=1$ and homogeneous for all other values of $\ell$.
\\

A nontrivial homogeneous solution requires the determinant to vanish. Using Eq.~\eqref{eq:fgdef}, the determinant is
\begin{equation}
\bigl[g_{\ell}f'_{\ell}-g'_{\ell}f_{\ell}\bigr]_{R}
=kR^{2}\Bigl[h^{(1)}_{\ell}(kR)\,j_{\ell}'(kR)-h^{(1)\prime}_{\ell}(kR)\,j_{\ell}(kR)\Bigr].
\label{eq:detexpand}
\end{equation}
The expression in brackets is a Wronskian. With $h^{(1)}_{\ell}=j_{\ell}+iy_{\ell}$, the $j_{\ell}j_{\ell}'$ terms cancel and leave $i\bigl(y_{\ell}j_{\ell}'-y_{\ell}'j_{\ell}\bigr)$. The standard spherical-Bessel Wronskian~\cite{dlmf}, $j_{\ell}(x)y_{\ell}'(x)-j_{\ell}'(x)y_{\ell}(x)=1/x^{2}$, then gives
\begin{equation}
\bigl[g_{\ell}f'_{\ell}-g'_{\ell}f_{\ell}\bigr]_{R}=kR^{2}\cdot\frac{-i}{(kR)^{2}}=-\frac{i}{k} ,
\label{eq:wronskian}
\end{equation}
the same nonzero determinant for every $\ell$. Each system with $\ell\neq1$ in Eq.~\eqref{eq:system} is therefore homogeneous and nonsingular, so
\begin{equation}
A_{\ell}=C_{\ell}=0
\qquad (\ell\neq1).
\label{eq:vanish}
\end{equation}
Both sums in Eq.~\eqref{eq:generalsol} reduce to a single term. This follows from the boundary conditions and holds at every frequency.

\subsection{Surviving Coefficient}

For $\ell=1$, Eq.~\eqref{eq:system} is inhomogeneous. Eliminating $A_{1}$ between its two rows gives
\begin{equation}
C_{1}\bigl[g_{1}f'_{1}-g'_{1}f_{1}\bigr]_{R}
=\frac{CR^{2}}{2}f'_{1}(R)-CR\,f_{1}(R).
\label{eq:eliminate}
\end{equation}
With $f_{1}(r)=r\,j_{1}(kr)$, the surface values are $f_{1}(R)=R\,j_{1}(x)$ and $f_{1}'(R)=j_{1}(x)+x\,j_{1}'(x)$, where $x=kR$. The right side becomes
\begin{equation}
\frac{CR^{2}}{2}\Bigl[j_{1}(x)+x\,j_{1}'(x)\Bigr]-CR^{2}j_{1}(x)
=\frac{CR^{2}}{2}\Bigl[x\,j_{1}'(x)-j_{1}(x)\Bigr].
\label{eq:rhsreduce}
\end{equation}
A single Bessel identity reduces the bracket in Eq.~\eqref{eq:rhsreduce}. Use $j_{1}'=j_{0}-2j_{1}/x$ followed by $j_{2}=3j_{1}/x-j_{0}$,
\begin{equation}
x\,j_{1}'(x)-j_{1}(x)=x\,j_{0}(x)-3j_{1}(x)=-x\,j_{2}(x).
\label{eq:besselid}
\end{equation}
Equation~\eqref{eq:besselid} introduces $j_{2}$ directly through the matching. The required combination is $xj_{1}'-j_{1}$, which is exactly $-xj_{2}$. The order-two function therefore comes from the boundary matching rather than from a later combination of separate contributions.
\\

\noindent Using Eqs.~\eqref{eq:wronskian} and~\eqref{eq:besselid} in Eq.~\eqref{eq:eliminate}, then restoring $C=-i\mu_{0}\omega P_{0}$, gives
\begin{equation}
C_{1}=-\frac{\mu_{0}\omega P_{0}k^{2}R^{3}}{2}\,j_{2}(kR).
\label{eq:C1}
\end{equation}
Together with Eq.~\eqref{eq:vanish}, this leaves a single term for the exterior magnetic field,
\begin{equation}
B_{\phi}^{\mathrm{out}}(r,\theta)=C_{1}\,h^{(1)}_{1}(kr)\,P^{1}_{1}(\cos\theta)
=\frac{\mu_{0}\omega P_{0}k^{2}R^{3}}{2}\,j_{2}(kR)\,h^{(1)}_{1}(kr)\,\sin\theta ,
\label{eq:Bout}
\end{equation}
which is exact at every exterior radius and frequency. Section~\ref{sec:exact} uses Eq.~\eqref{eq:Bout} to obtain the radiation field and power.
\\

\emph{Exterior Zeros.} Every component of the exterior field is proportional to the single coefficient $C_{1}$. The field therefore vanishes everywhere outside the sphere if and only if $C_{1}=0$, which by Eq.~\eqref{eq:C1} occurs if and only if
\begin{equation}
j_{2}(kR)=0 .
\label{eq:zeros}
\end{equation}
An exterior uniqueness theorem is unnecessary here. The field contains one coefficient, and one condition sets it.

\section{Exact Field, Power, and the Bare-Sphere Comparison}
\label{sec:exact}

\subsection{Interior Coefficient}
\label{sec:interior}

Equation~\eqref{eq:eliminate} gives the exterior coefficient after $A_{1}$ is eliminated from Eq.~\eqref{eq:system}. The interior coefficient follows by multiplying the first row of Eq.~\eqref{eq:system} by $g_{1}'(R)$ and the second by $g_{1}(R)$. Their difference eliminates $C_{1}$ and gives
\begin{equation}
A_{1}\bigl[g_{1}f_{1}'-g_{1}'f_{1}\bigr]_{R}
=\frac{CR^{2}}{2}\,g_{1}'(R)-CR\,g_{1}(R).
\label{eq:eliminateA}
\end{equation}
The left side contains the Wronskian already evaluated in Eq.~\eqref{eq:wronskian}. For the right side, use $g_{1}(r)=r\,h^{(1)}_{1}(kr)$ from Eq.~\eqref{eq:fgdef}. Then $g_{1}(R)=R\,h^{(1)}_{1}(x)$ and $g_{1}'(R)=h^{(1)}_{1}(x)+x\,h^{(1)\prime}_{1}(x)$, so
\begin{equation}
\frac{CR^{2}}{2}\Bigl[h^{(1)}_{1}(x)+x\,h^{(1)\prime}_{1}(x)\Bigr]-CR^{2}\,h^{(1)}_{1}(x)
=\frac{CR^{2}}{2}\Bigl[x\,h^{(1)\prime}_{1}(x)-h^{(1)}_{1}(x)\Bigr],
\label{eq:rhsreduceA}
\end{equation}
which has the same form as Eq.~\eqref{eq:rhsreduce}, with $h^{(1)}_{1}$ replacing $j_{1}$. The derivative and recurrence relations used in Eq.~\eqref{eq:besselid} connect orders one and two within a single spherical-Bessel family. They hold for $j_{\ell}$ and $y_{\ell}$, and therefore for any fixed linear combination of the two~\cite{dlmf}. In particular, they hold for $h^{(1)}_{\ell}=j_{\ell}+iy_{\ell}$,
\begin{equation}
x\,h^{(1)\prime}_{1}(x)-h^{(1)}_{1}(x)=-\,x\,h^{(1)}_{2}(x).
\label{eq:besselidh}
\end{equation}
Substitution of Eqs.~\eqref{eq:wronskian} and~\eqref{eq:besselidh} into Eq.~\eqref{eq:eliminateA}, followed by $C=-i\mu_{0}\omega P_{0}$, gives
\begin{equation}
A_{1}=-\frac{\mu_{0}\omega P_{0}k^{2}R^{3}}{2}\,h^{(1)}_{2}(kR).
\label{eq:A1}
\end{equation}
The interior and exterior coefficients can now be written together. Using $P^{1}_{1}=-\sin\theta$ as in Eq.~\eqref{eq:Bout}, both fields share one overall constant,
\begin{equation}
\begin{aligned}
B_{\phi}^{\mathrm{in}}&=\frac{\mu_{0}\omega P_{0}k^{2}R^{3}}{2}\,h^{(1)}_{2}(kR)\,j_{1}(kr)\,\sin\theta,\\
B_{\phi}^{\mathrm{out}}&=\frac{\mu_{0}\omega P_{0}k^{2}R^{3}}{2}\,j_{2}(kR)\,h^{(1)}_{1}(kr)\,\sin\theta.
\end{aligned}
\label{eq:Bpair}
\end{equation}
The order-one function in each region controls radial propagation. Regularity requires $j_{1}(kr)$ inside, while outgoing behavior requires $h^{(1)}_{1}(kr)$ outside. The surface factors have order two, with $h^{(1)}_{2}(kR)$ multiplying the interior field and $j_{2}(kR)$ multiplying the exterior field. The matching exchanges the roles of $j$ and $h$. The pair also reproduces the tangential jump in Eq.~\eqref{eq:bcBtan}. Subtract the two fields in Eq.~\eqref{eq:Bpair} at $r=R$ and use $j_{2}h^{(1)}_{1}-h^{(1)}_{2}j_{1}=i\,(j_{2}y_{1}-y_{2}j_{1})$ together with the cross Wronskian $j_{2}(x)\,y_{1}(x)-j_{1}(x)\,y_{2}(x)=1/x^{2}$~\cite{dlmf},
\begin{equation}
B_{\phi}^{\mathrm{out}}-B_{\phi}^{\mathrm{in}}\Big|_{r=R}
=\frac{\mu_{0}\omega P_{0}k^{2}R^{3}}{2}\,\frac{i}{x^{2}}\,\sin\theta
=\frac{i\mu_{0}\omega P_{0}R}{2}\,\sin\theta,
\label{eq:crossjump}
\end{equation}
which equals $-\mu_{0}K_{\theta}$ for the phasor form of the sheet current in Eq.~\eqref{eq:Kf}, as required by Eq.~\eqref{eq:bcBtan}.
\\

The exterior coefficient vanishes at every root of $j_{2}$, but the interior coefficient never vanishes at a real frequency. Away from the roots of $j_{2}$, the real part of $h^{(1)}_{2}=j_{2}+iy_{2}$ is already nonzero. At a positive root $x_{0}$ of $j_{2}$, the order-two Wronskian used before Eq.~\eqref{eq:wronskian}, evaluated at $x_{0}$, reduces to
\begin{equation}
-\,j_{2}'(x_{0})\,y_{2}(x_{0})=\frac{1}{x_{0}^{2}},
\label{eq:noshare}
\end{equation}
so $y_{2}(x_{0})\neq0$ and $h^{(1)}_{2}(x_{0})=i\,y_{2}(x_{0})\neq0$. The interior field survives at every nonradiating frequency, where its coefficient becomes purely imaginary. Section~\ref{sec:ledger} connects this phase to the energy balance.

\subsection{Exact Exterior Field}
\label{sec:exterior}

The complete exterior field follows from Eq.~\eqref{eq:Bout}. The electric field comes from Eq.~\eqref{eq:Efromcurl} with $\vJ_{\omega}=\zero$ in $r>R$, so $\vE=(ic^{2}/\omega)\,\nabla\times\vB$. The needed curl components are given in Eq.~\eqref{eq:curlB}. With $B_{\phi}=C_{1}h^{(1)}_{1}(kr)\,P^{1}_{1}(\cos\theta)$ and $P^{1}_{1}=-\sin\theta$, the radial component follows from $\partial_{\theta}\bigl(\sin\theta\,B_{\phi}\bigr)=-2C_{1}h^{(1)}_{1}(kr)\sin\theta\cos\theta$, which gives
\begin{equation}
E_{r}^{\mathrm{out}}=-\frac{2ic^{2}C_{1}}{\omega r}\,h^{(1)}_{1}(kr)\,\cos\theta.
\label{eq:Erout}
\end{equation}
For the tangential component, $\partial_{r}\bigl(rB_{\phi}\bigr)=-C_{1}\sin\theta\bigl[h^{(1)}_{1}(kr)+kr\,h^{(1)\prime}_{1}(kr)\bigr]$. The identity $h^{(1)\prime}_{1}=h^{(1)}_{0}-2h^{(1)}_{1}/x$, valid for the Hankel family by the argument leading to Eq.~\eqref{eq:besselidh}, changes the bracket to $kr\,h^{(1)}_{0}(kr)-h^{(1)}_{1}(kr)$. Hence
\begin{equation}
E_{\theta}^{\mathrm{out}}=\frac{ic^{2}C_{1}}{\omega r}\Bigl[kr\,h^{(1)}_{0}(kr)-h^{(1)}_{1}(kr)\Bigr]\sin\theta.
\label{eq:Ethetaout}
\end{equation}
The low-order spherical Bessel and Hankel functions have the closed forms~\cite{dlmf}
\begin{equation}
j_0(x)=\frac{\sin x}{x},\qquad
j_1(x)=\frac{\sin x}{x^{2}}-\frac{\cos x}{x},\qquad
j_2(x)=\Bigl(\frac{3}{x^{3}}-\frac{1}{x}\Bigr)\sin x-\frac{3\cos x}{x^{2}},
\label{eq:jdefs}
\end{equation}
\begin{equation}
h^{(1)}_{0}(x)=-\frac{i\,e^{ix}}{x},
\qquad
h^{(1)}_{1}(x)=-\frac{e^{ix}}{x}\Bigl(1+\frac{i}{x}\Bigr).
\label{eq:hclosed}
\end{equation}
Every term in Eqs.~\eqref{eq:Erout} and~\eqref{eq:Ethetaout} is an outgoing factor $e^{ikr}$ multiplied by a polynomial in $1/(kr)$. The tangential field contains the radiative $1/r$ term along with $1/r^{2}$ and $1/r^{3}$ near-field terms. The radial field begins at $1/r^{2}$ and has no radiation-zone contribution.
\\

The exterior field is exactly that of an ideal point electric dipole. Both this solution and the field of a point dipole $p\,e^{-i\omega t}\zhat$ at the origin~\cite{jackson} satisfy the source-free Maxwell equations in $r>R$, obey outgoing conditions, and contain only the electric-parity $\ell=1$ multipole. Their radial dependence is therefore fixed up to one constant. Matching any field component gives
\begin{equation}
p_{b}=\frac{4\pi}{3}R^{3}P_{0},
\qquad
p_{\mathrm{eff}}=-\frac{4\pi C_{1}}{\mu_{0}ck^{3}}
=2\pi P_{0}R^{3}\,j_{2}(kR)=\frac{3}{2}\,j_{2}(kR)\,p_{b},
\label{eq:peff}
\end{equation}
where $p_{b}$ is the electric dipole moment of the sphere without the coating. At every exterior radius, not only in the radiation zone, the neutralized sphere with the minimal sheet current has the same field as a point dipole with moment $\tfrac32\,j_{2}(kR)\,p_{b}$. This real factor can change sign. A negative value produces a phase shift of $\pi$ in every field component. At a root of $j_{2}$, the effective moment and the entire exterior field vanish, in agreement with Eq.~\eqref{eq:zeros}.

\subsection{Radiation Zone and the Radiated Power}
\label{sec:radzone}

In the radiation zone, $kr\gg1$. Equation~\eqref{eq:hclosed} gives $h^{(1)}_{1}(kr)\rightarrow-e^{ikr}/(kr)$, consistent with Eq.~\eqref{eq:hankelasym}, and the exact identity $kr\,h^{(1)}_{0}(kr)=-ie^{ikr}$. Equation~\eqref{eq:Ethetaout} then retains only one $1/r$ term,
\begin{equation}
E_{\theta}^{\mathrm{out}}\longrightarrow
\frac{ic^{2}C_{1}}{\omega r}\bigl(-ie^{ikr}\bigr)\sin\theta
=\frac{c^{2}C_{1}}{\omega r}\,e^{ikr}\sin\theta
=-\frac{\mu_{0}\omega^{2}R^{3}P_{0}}{2r}\,j_{2}(kR)\,e^{ikr}\sin\theta,
\label{eq:Eradcomplex}
\end{equation}
where Eq.~\eqref{eq:C1} and $c^{2}k^{2}=\omega^{2}$ were used in the final step. Restoring $e^{-i\omega t}$ and taking the real part gives
\begin{equation}
\vE_{\mathrm{rad}}(\vx,t)=
-\frac{\mu_0\omega^2R^3P_0}{2r}\,j_2(kR)
\cos\!\left[\omega\left(t-\frac{r}{c}\right)\right]
\sin\theta\,\thetahat+O(r^{-2}),
\label{eq:Erad}
\end{equation}
where $\theta$ is measured from the polarization axis. The $O(r^{-2})$ term contains the near-field contributions from Eqs.~\eqref{eq:Erout} and~\eqref{eq:Ethetaout}. The same limit applied to Eq.~\eqref{eq:Bout} gives $B_{\phi}=E_{\theta}/c$, and therefore
\begin{equation}
\vB_{\mathrm{rad}}=\frac{1}{c}\nhat\times\vE_{\mathrm{rad}}+O(r^{-2}).
\label{eq:Brad}
\end{equation}
Equation~\eqref{eq:Erad} is exact in $kR$. Its angular factor is the familiar $\sin\theta$ pattern of an electric-parity $\ell=1$ field, while its dependence on frequency and source size differs from an ordinary electric dipole. The time-averaged Poynting flux is $\langle\vS\rangle=\langle E_{\mathrm{rad}}^2\rangle\nhat/(\mu_0c)$. Using Eq.~\eqref{eq:Erad} and $\langle\cos^{2}\rangle=1/2$ gives the angular distribution
\begin{equation}
\frac{\dd\overline{\mathcal P}}{\dd\Omega}
=r^{2}\,\langle\vS\rangle\cdot\nhat
=\frac{\mu_0\omega^{4}R^{6}P_0^{2}}{8c}\,j_2^{2}(kR)\,\sin^{2}\theta .
\label{eq:dPdOmega}
\end{equation}
The solid-angle integral follows by setting $u=\cos\theta$,
\begin{equation}
\int\sin^{2}\theta\dd\Omega
=2\pi\int_{0}^{\pi}\sin^{3}\theta\dd\theta
=2\pi\int_{-1}^{1}\bigl(1-u^{2}\bigr)\dd u
=\frac{8\pi}{3},
\label{eq:solidangle}
\end{equation}
so the total radiated power is
\begin{equation}
\overline{\mathcal P}=
\frac{\pi\mu_0\omega^4R^6P_0^2}{3c}\,j_2^2(kR).
\label{eq:power}
\end{equation}
The units are those of power because $P_0R^3$ has the dimensions of an electric dipole moment. Both the field and the power vanish as $\omega\rightarrow0$.
\\

An ideal point dipole with the bare moment $p_b$ from Eq.~\eqref{eq:peff} provides the reference
\begin{equation}
E_{\mathrm{dip}}=\frac{\mu_0\omega^{2}p_b}{4\pi r}\sin\theta,
\qquad
\overline{\mathcal P}_{\mathrm{dip}}=\frac{\mu_0\omega^{4}p_b^{2}}{12\pi c}.
\label{eq:dipref}
\end{equation}
At fixed frequency and angle, let $E_{\mathrm{sheet}}$ and $E_{\mathrm{dip}}$ be the signed coefficients of the shared pattern $-\cos[\omega(t-r/c)]\,\sin\theta\,\thetahat$ in Eq.~\eqref{eq:Erad} and in the standard point-dipole field. Both are positive for small $kR$. The field and power ratios are
\begin{equation}
\frac{E_{\mathrm{sheet}}}{E_{\mathrm{dip}}}
=\frac{3}{2}j_2(kR),
\qquad
\frac{\overline{\mathcal P}_{\mathrm{sheet}}}
{\overline{\mathcal P}_{\mathrm{dip}}}
=\frac{9}{4}j_2^2(kR).
\label{eq:dipoleratios}
\end{equation}
Equation~\eqref{eq:peff} extends the same comparison to the full exterior field. A negative field ratio means a phase reversal of $\pi$, not negative power. The point dipole serves only as a reference because the neutralized source itself has zero electric dipole moment.

\subsection{Energy Account}
\label{sec:ledger}

The current in Eq.~\eqref{eq:Jtot} is impressed and must be maintained by an external agent against its self-field. In steady state, the average supplied power leaves as radiation because the fields are periodic and the model contains no dissipation. Poynting's theorem states this balance locally~\cite{zangwill}. The field transfers energy to the sources at the rate $\vJ\cdot\vE$ per unit volume, so the agent supplies the negative of this quantity. The required cycle average is therefore $-\int\vJ_{\mathrm{tot}}\cdot\vE\dd^{3}x$. For two phasors with the common time factor $e^{-i\omega t}$, the cycle average of the product of their real parts is
\begin{equation}
\Bigl\langle\,\mathrm{Re}\bigl[a\,e^{-i\omega t}\bigr]\,
\mathrm{Re}\bigl[b\,e^{-i\omega t}\bigr]\Bigr\rangle
=\tfrac12\,\mathrm{Re}\bigl(a\,b^{*}\bigr),
\label{eq:avgrule}
\end{equation}
so the average input power is
\begin{equation}
\overline{\mathcal P}_{\mathrm{in}}
=-\frac12\,\mathrm{Re}\int\vE\cdot\vJ_{\omega}^{*}\dd^{3}x,
\label{eq:Pin}
\end{equation}
where the integral includes both the ball and the coating. Since $\sigma_{\mathrm{tot}}=0$, the two conditions in Eq.~\eqref{eq:bcE} make $\vE$ continuous across $r=R$. The field acting on the sheet current is therefore unambiguous.
\\

The calculation requires the interior electric field. Inside the sphere, where $\vJ_{\omega}=-i\omega P_{0}\zhat$, Eq.~\eqref{eq:Efromcurl} separates it into a uniform term and an induced term,
\begin{equation}
\vE_{\mathrm{in}}=-\frac{P_{0}}{\varepsilon_{0}}\,\zhat
+\frac{ic^{2}}{\omega}\,\nabla\times\vB_{\mathrm{in}},
\label{eq:Einsplit}
\end{equation}
Applying the curl components in Eq.~\eqref{eq:curlB} to the interior field in Eq.~\eqref{eq:Bpair}, with $\zhat=\cos\theta\,\rhat-\sin\theta\,\thetahat$ and the derivative identity used in Eq.~\eqref{eq:Ethetaout}, gives
\begin{equation}
E_{r}^{\mathrm{in}}=-\Bigl[\frac{P_{0}}{\varepsilon_{0}}
+\frac{2ic^{2}A_{1}}{\omega r}\,j_{1}(kr)\Bigr]\cos\theta,
\qquad
E_{\theta}^{\mathrm{in}}=\Bigl[\frac{P_{0}}{\varepsilon_{0}}
+\frac{ic^{2}A_{1}}{\omega r}\bigl(kr\,j_{0}(kr)-j_{1}(kr)\bigr)\Bigr]\sin\theta.
\label{eq:Ein}
\end{equation}
The limit $\omega\rightarrow0$ checks the result. Since $h^{(1)}_{2}(kR)\rightarrow-3i/(kR)^{3}$, the coefficient tends to $A_{1}\rightarrow\tfrac32 i\mu_{0}cP_{0}$. The interior magnetic field in Eq.~\eqref{eq:Bpair} then approaches the quasistatic Amp\`ere field $\tfrac12\mu_{0}(\dd P_{s}/\dd t)\,r\sin\theta\,\phihat$ of the uniform interior current. The two terms of order unity in Eq.~\eqref{eq:Ein} cancel, leaving an electric field of order $(kR)^{2}$. This is consistent with a source whose charge vanishes identically and whose current goes to zero with $\omega$. The same magnetic field follows from $\nabla\times\vH=\partial\vD/\partial t$. There is no free interior current, and Sec.~\ref{sec:statics} gives $\vD=\vP$ at leading order.
\\

The volume part of Eq.~\eqref{eq:Pin} can now be evaluated. With $\vJ_{\omega}^{*}=+i\omega P_{0}\zhat$ and $\zhat\cdot\vE=E_{r}\cos\theta-E_{\theta}\sin\theta$, the angular integrals are $\int\cos^{2}\theta\dd\Omega=4\pi/3$ and $\int\sin^{2}\theta\dd\Omega=8\pi/3$. Multiplication of the uniform term in Eq.~\eqref{eq:Ein} by $i\omega P_{0}$ produces a purely imaginary quantity, which drops out of the real part in Eq.~\eqref{eq:Pin}. A field in phase with $\vP$ exchanges no average energy with a current in phase with $\dd\vP/\dd t$. The remaining terms are proportional to $A_{1}$. Since $\tfrac{4\pi}{3}\cdot2\,j_{1}+\tfrac{8\pi}{3}\bigl(kr\,j_{0}-j_{1}\bigr)=\tfrac{8\pi}{3}\,kr\,j_{0}$, the angular contributions reduce to one radial integral,
\begin{equation}
\int_{r<R}\vE\cdot\vJ_{\omega}^{*}\dd^{3}x
=\frac{8\pi}{3}\,c^{2}P_{0}A_{1}\,k\int_{0}^{R}j_{0}(kr)\,r^{2}\dd r
+\text{(imag.)},
\label{eq:volassembled}
\end{equation}
where (imag.) denotes purely imaginary terms that do not contribute to Eq.~\eqref{eq:Pin}. The radial integral follows from the derivative identity $\dd\bigl[u^{2}j_{1}(u)\bigr]/\dd u=u^{2}j_{0}(u)$, which is the same relation used at Eq.~\eqref{eq:besselid}. With $u=kr$,
\begin{equation}
k\int_{0}^{R}j_{0}(kr)\,r^{2}\dd r
=\frac{1}{k^{2}}\int_{0}^{x}u^{2}j_{0}(u)\dd u
=\frac{x^{2}}{k^{2}}\,j_{1}(x)=R^{2}j_{1}(x).
\label{eq:radialint}
\end{equation}
The coating contributes through its tangential current. From Eq.~\eqref{eq:Kf}, $\vK_{\omega}^{*}=+i\omega P_{0}(R/2)\sin\theta\,\thetahat$. Evaluating $E_{\theta}$ from Eq.~\eqref{eq:Ein} at $r=R$ gives
\begin{equation}
\oint\vE\cdot\vK_{\omega}^{*}\dd S
=-\frac{4\pi}{3}\,c^{2}P_{0}A_{1}R^{2}\bigl[x\,j_{0}(x)-j_{1}(x)\bigr]
+\text{(imag.)}.
\label{eq:sheetwork}
\end{equation}
Combining Eqs.~\eqref{eq:volassembled} and~\eqref{eq:sheetwork} with the radial integral in Eq.~\eqref{eq:radialint} gives
\begin{equation}
\begin{aligned}
\int\vE\cdot\vJ_{\omega}^{*}\dd^{3}x
&=\frac{4\pi}{3}\,c^{2}P_{0}A_{1}R^{2}\bigl[3j_{1}(x)-x\,j_{0}(x)\bigr]+\text{(imag.)}\\
&=\frac{4\pi}{3}\,c^{2}P_{0}A_{1}R^{2}\,x\,j_{2}(x)+\text{(imag.)},
\end{aligned}
\label{eq:worktotal}
\end{equation}
because Eq.~\eqref{eq:besselid} can be read as $3j_{1}-xj_{0}=xj_{2}$. The same identity that introduces $j_{2}$ in the boundary matching appears again in the energy calculation. Only the real part of $A_{1}$ contributes to Eq.~\eqref{eq:Pin}. Writing $h^{(1)}_{2}=j_{2}+iy_{2}$ in Eq.~\eqref{eq:A1} gives $\mathrm{Re}\,A_{1}=-(\mu_{0}\omega P_{0}k^{2}R^{3}/2)\,j_{2}(x)$, and therefore
\begin{equation}
\overline{\mathcal P}_{\mathrm{in}}
=-\frac{2\pi}{3}\,c^{2}P_{0}R^{2}\,x\,j_{2}(x)\,\mathrm{Re}\,A_{1}
=\frac{\pi\mu_{0}\omega^{4}R^{6}P_{0}^{2}}{3c}\,j_{2}^{2}(kR),
\label{eq:Pinresult}
\end{equation}
which is identical to Eq.~\eqref{eq:power}. The average power supplied by the agent driving the polarization and coating is exactly the power carried away by radiation. The imaginary terms describe energy stored and returned during each cycle. This balance holds across any sphere with radius $r>R$. Integrating the time-averaged radial Poynting flux from the exact fields in Eqs.~\eqref{eq:Bout} and~\eqref{eq:Ethetaout} gives Eq.~\eqref{eq:power} independently of $r$. The $1/r^{2}$ and $1/r^{3}$ terms contribute only reactive energy.
\\

At a root $x_{0}$ of $j_{2}$, the real part of $A_{1}$ vanishes although $A_{1}$ remains nonzero,
\begin{equation}
A_{1}=-\,i\,\frac{\mu_{0}\omega P_{0}k^{2}R^{3}}{2}\,y_{2}(x_{0})\neq0,
\label{eq:A1root}
\end{equation}
as follows from Eq.~\eqref{eq:noshare}. Currents and interior fields therefore remain. When $A_{1}$ is purely imaginary, each bracket in Eq.~\eqref{eq:Ein} is real. Every interior field component is then in phase or out of phase by $\pi$ with the polarization and has no quadrature component. The current, being proportional to the time derivative of the polarization, remains in quadrature. Their product averages to zero over a cycle. The source is nonradiating and requires no net work, although energy is stored and recovered within each period.

\subsection{Comparison With the Bare Sphere}
\label{sec:bare}

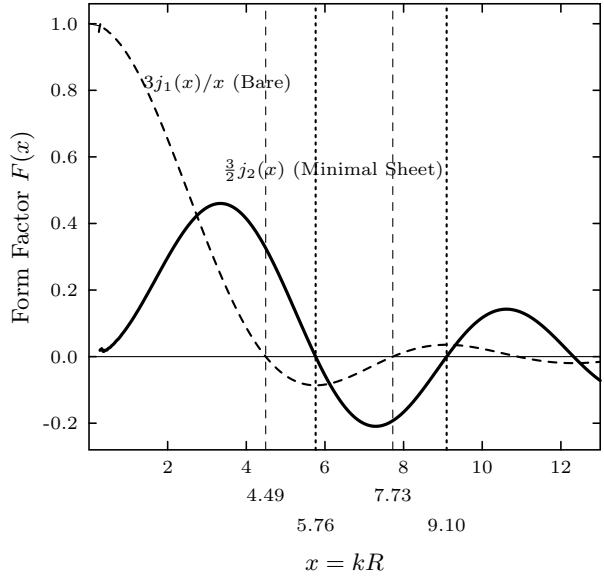
\begin{figure}[!htbp]
\centering
\begin{tikzpicture}[line cap=round,x=0.52cm,y=4.4cm]
  \draw[semithick] (0,-0.28) rectangle (13,1.06);
  \foreach \x in {2,4,6,8,10,12}
    {\draw[semithick] (\x,-0.28) -- (\x,-0.255);
     \draw[semithick] (\x,1.06) -- (\x,1.035);
     \node[below] at (\x,-0.285) {\footnotesize $\x$};}
  \foreach \y/\yl in {-0.2/-0.2,0/0.0,0.2/0.2,0.4/0.4,0.6/0.6,0.8/0.8,1/1.0}
    {\draw[semithick] (0,\y) -- (0.18,\y);
     \draw[semithick] (13,\y) -- (12.82,\y);
     \node[left] at (-0.05,\y) {\footnotesize \yl};}
  \draw[thin] (0,0) -- (13,0);
  \draw (0,1) -- (0.25,0.994);
  \draw[thick,dashed,domain=0.25:13,samples=300,smooth]
    plot (\x,{3*sin(deg(\x))/(\x*\x*\x)-3*cos(deg(\x))/(\x*\x)});
  \draw[very thick,domain=0.28:13,samples=300,smooth]
    plot (\x,{1.5*((3/(\x*\x*\x)-1/\x)*sin(deg(\x))-3*cos(deg(\x))/(\x*\x))});
  \draw[thin,dashed] (4.4934,-0.28) -- (4.4934,1.06);
  \draw[thin,dashed] (7.7253,-0.28) -- (7.7253,1.06);
  \draw[dotted,thick] (5.7635,-0.28) -- (5.7635,1.06);
  \draw[dotted,thick] (9.0950,-0.28) -- (9.0950,1.06);
  \node at (4.4934,-0.415) {\footnotesize $4.49$};
  \node at (7.7253,-0.415) {\footnotesize $7.73$};
  \node at (5.7635,-0.505) {\footnotesize $5.76$};
  \node at (9.0950,-0.505) {\footnotesize $9.10$};
  \node[anchor=west] at (1.15,0.82) {\footnotesize $3j_1(x)/x$ (Bare)};
  \node[anchor=west] at (3.15,0.56) {\footnotesize $\tfrac32 j_2(x)$ (Minimal Sheet)};
  \node at (6.5,-0.615) {\small $x=kR$};
  \node[rotate=90,anchor=south] at (-1.15,0.39) {\small Form Factor $F(x)$};
\end{tikzpicture}
\caption{The radiating form factors in Eq.~\eqref{eq:formfactors}. The dashed curve shows the bare-sphere factor $3j_1(x)/x$, which approaches the point-dipole limit $F=1$ as $x\rightarrow0$. It vanishes at the roots of $j_1$, including $x\simeq4.49$ and $7.73$, shown by thin dashed vertical lines. The solid curve shows the minimal tangential-sheet factor $\tfrac32 j_2(x)$, which approaches $x^{2}/10$ at long wavelength. It vanishes at the roots of $j_2$, including $x\simeq5.76$ and $9.10$, shown by dotted vertical lines. The two root families have no common positive value. The volume-current realization lies on the zero line.}
\label{fig:formfactors}
\end{figure}

The same boundary method applies to the bare sphere, whose exact fields were obtained by Mansuripur and Jakobsen~\cite{mansuripur2020}. Removing the coating removes the free surface charge and sheet current, while the bound surface charge $\sigma_b$ and volume current $\vJ_b$ remain. The second row of Eq.~\eqref{eq:system}, which follows from tangential $\vE$ and the interior current, is unchanged. The first row becomes homogeneous because the surface current is absent,
\begin{equation}
\begin{pmatrix} g_{1}(R) & -f_{1}(R)\\[2pt] g_{1}'(R) & -f_{1}'(R)\end{pmatrix}
\begin{pmatrix} C_{1}^{\mathrm{bare}}\\[2pt] A_{1}^{\mathrm{bare}}\end{pmatrix}
=\begin{pmatrix} 0\\[2pt] CR\end{pmatrix}.
\label{eq:baresystem}
\end{equation}
The normal component of $\vE$ now jumps by $\sigma_{b}/\varepsilon_{0}$. Evaluation with the interior $J_{r}$ retained gives the same first row, so the bare-sphere system also remains two by two. Eliminating $A_{1}^{\mathrm{bare}}$ as in Eq.~\eqref{eq:eliminate} requires multiplication of the first row by $f_{1}'(R)$ and the second by $f_{1}(R)$. Subtraction then gives
\begin{equation}
C_{1}^{\mathrm{bare}}\bigl[g_{1}f_{1}'-g_{1}'f_{1}\bigr]_{R}
=-CR\,f_{1}(R)=-CR^{2}j_{1}(x),
\label{eq:bareeliminate}
\end{equation}
and the Wronskian in Eq.~\eqref{eq:wronskian} then gives
\begin{equation}
C_{1}^{\mathrm{bare}}=-\,\mu_{0}\omega P_{0}kR^{2}\,j_{1}(kR).
\label{eq:C1bare}
\end{equation}
The argument in Sec.~\ref{sec:exterior} shows that the bare sphere also has the exact exterior field of a point dipole, now with
\begin{equation}
p_{\mathrm{eff}}^{\mathrm{bare}}=-\frac{4\pi C_{1}^{\mathrm{bare}}}{\mu_{0}ck^{3}}
=\frac{4\pi P_{0}R^{3}}{x}\,j_{1}(x)=\frac{3j_{1}(x)}{x}\,p_{b}.
\label{eq:peffbare}
\end{equation}
As $\omega\rightarrow0$, the bare solution approaches the static fields in Sec.~\ref{sec:statics}. The interior electric field tends to $-P_{s}\,\zhat/(3\varepsilon_{0})$, and the exterior field tends to the static dipole field of $p_{b}$. For either radiating source, the amplitude relative to the point-dipole reference is $p_{\mathrm{eff}}/p_{b}$. Therefore
\begin{equation}
F_{\mathrm{bare}}(x)=\frac{3j_1(x)}{x},
\qquad
F_{\mathrm{sheet}}(x)=\frac{3}{2}\,j_2(x),
\qquad x=kR.
\label{eq:formfactors}
\end{equation}
For the volume-current realization, $F=0$ at every frequency. The long-wavelength expansions are
\begin{equation}
\frac{3j_1(x)}{x}=1-\frac{x^{2}}{10}+\frac{x^{4}}{280}+O(x^{6}),
\qquad
\frac{3}{2}\,j_2(x)=\frac{x^{2}}{10}-\frac{x^{4}}{140}+O(x^{6}).
\label{eq:series}
\end{equation}
The bare sphere approaches ordinary point-dipole radiation as $x\rightarrow0$. The minimal sheet realization has no leading dipole term and begins two powers of $x$ later. Complete cancellation of the current removes radiation at every frequency. The bound and free subsystems are separately conserved, so each has its own form factor. The bound subsystem is the bare sphere and has $F_b(x)=3j_1(x)/x$. By linearity, the free subsystem made from the neutralizing surface charge and tangential current has $F_f(x)=F_{\mathrm{sheet}}(x)-F_b(x)$. Using the recurrence $3j_{1}(x)/x=j_{0}(x)+j_{2}(x)$ from Eq.~\eqref{eq:besselid}, this difference becomes
\begin{equation}
F_{f}(x)=\frac{3}{2}\,j_{2}(x)-\frac{3j_{1}(x)}{x}
=\frac{j_{2}(x)-2j_{0}(x)}{2}\,.
\label{eq:freeclosed}
\end{equation}
Its long-wavelength expansion is
\begin{equation}
F_f(x)=-1+\frac{x^{2}}{5}-\frac{3x^{4}}{280}+O(x^{6})
\label{eq:freeff}
\end{equation}
and begins at $-1$, as Eq.~\eqref{eq:freeclosed} also shows because $j_{0}(0)=1$ and $j_{2}(0)=0$. The leading radiation of the free subsystem cancels the point-dipole term of the bound subsystem in Eq.~\eqref{eq:series}. Finite-size retardation leaves a residual two powers of $kR$ higher. That residual also vanishes at the roots of $j_2$. The free subsystem alone is silent when $j_{2}(x)=2j_{0}(x)$, which defines a separate root sequence. In the volume-current realization, the free subsystem instead has the exact form factor $-F_b(x)$ and cancels the bound radiation at every order and frequency. Away from roots of $j_1$, the amplitude ratio of the two radiating complete sources is
\begin{equation}
\frac{E_{\mathrm{sheet}}}{E_{\mathrm{bare}}}
=\frac{x\,j_2(x)}{2\,j_1(x)}.
\label{eq:exactratio}
\end{equation}
The finite bare sphere is exactly nonradiating at each positive root of $j_1$,
\begin{equation}
x\simeq4.493409,\quad 7.725252,\quad 10.904122,\ldots.
\label{eq:j1zeros}
\end{equation}
The minimal sheet realization changes its form factor from $3j_1(x)/x$ to $\tfrac32 j_2(x)$. It therefore shifts the finite-size silence sequence from the roots of $j_1$ to the roots of $j_2$. The volume-current realization supports the same charge history but has no radiating channel at any frequency. Table~\ref{tab:realizations} and Fig.~\ref{fig:formfactors} compare the three cases.

\begin{table}[!htbp]
\caption{Three source realizations based on the same oscillating uniform polarization, with $x=kR$. The form factor $F(x)$ is the radiation amplitude relative to the point-dipole reference in Eq.~\eqref{eq:dipref}. The first roots of $j_1$ are $x\simeq4.49$ and $7.73$, while the first roots of $j_2$ are $x\simeq5.76$ and $9.10$. The bare sphere, solved exactly in Ref.~\cite{mansuripur2020}, is included as a reference. Only the two neutralized realizations have the same prescribed charge history.}
\label{tab:realizations}
\centering\small
\begin{tabularx}{\textwidth}{@{}L{0.25\textwidth}L{0.13\textwidth}L{0.21\textwidth}L{0.12\textwidth}Y@{}}
\toprule
Realization & $\rho_{\mathrm{tot}}$ & $\vJ_{\mathrm{tot}}$ & $F(x)$ & Nonradiating at\\
\midrule
Bare Sphere & $\sigma_b\,\delta(r-R)$ & $\vJ_b\,\Theta(R-r)$ & $3j_1(x)/x$ & Roots of $j_1$\\
Neutralized, Volume Current & $0$ & $\mathbf 0$ & $0$ & All Frequencies\\
Neutralized, Minimal Sheet & $0$ & $\neq\mathbf 0$, Divergence-Free & $\tfrac32\,j_2(x)$ & Roots of $j_2$\\
\bottomrule
\end{tabularx}
\end{table}

\FloatBarrier

\section{Interpretation}
\label{sec:interpret}

\subsection{Why Zero Charge Multipoles Do Not Settle the Question}

Although $\rho_{\mathrm{tot}}=0$ as a distribution and every electric charge multipole vanishes, Eq.~\eqref{eq:Jtot} contains a nonzero current. The volume polarization current and tangential sheet current cancel each other's divergence without cancelling their transverse spatial structure. The boundary conditions in Sec.~\ref{sec:fields} respond to that transverse structure, which fixes the single nonzero coefficient in Eq.~\eqref{eq:C1}.
\\

The contrast between the two realizations in Sec.~\ref{sec:source} comes from the freedom left by surface-charge continuity, which imposes only a divergence constraint. The volume-current realization sets the entire total current to zero, while the sheet-current realization removes only its longitudinal source. A time-dependent surface charge therefore does not fully specify the electromagnetic source. Its transport path must also be stated. The same conclusion holds for any prescribed $\rho(\vx,t)$,
\begin{equation}
\nabla\cdot\vJ=-\frac{\partial\rho}{\partial t},
\qquad
\vJ\longrightarrow\vJ+\nabla\times\mathbf W
\quad\text{leaves }\rho\text{ unchanged for every }\mathbf W .
\label{eq:generalfreedom}
\end{equation}
Retarded fields do not remove this freedom. The time-dependent field expressions give $\vE$ and $\vB$ directly in terms of $\rho$, $\vJ$, and their retarded time derivatives~\cite{griffithsheald}. They select the outgoing solution for a specified source, not a unique current for a specified charge density.

\subsection{Long-Wavelength Suppression}

For $x\ll1$,
\\
\begin{equation}
 j_2(x)=\frac{x^2}{15}-\frac{x^4}{210}+O(x^6),
\label{eq:j2small}
\end{equation}
\\
which gives
\\
\begin{equation}
\vE_{\mathrm{rad}}\simeq
-\frac{\mu_0\omega^4R^5P_0}{30c^2r}
\cos\!\left[\omega\left(t-\frac{r}{c}\right)\right]
\sin\theta\,\thetahat
\label{eq:Esmall}
\end{equation}
\\
and
\\
\begin{equation}
\overline{\mathcal P}\simeq
\frac{\pi\mu_0\omega^8R^{10}P_0^2}{675c^5}.
\label{eq:Psmall}
\end{equation}
\\
Equation~\eqref{eq:dipoleratios} then gives the suppression directly,
\begin{equation}
\frac{E_{\mathrm{sheet}}}{E_{\mathrm{dip}}}=\frac{(kR)^{2}}{10}+O\bigl[(kR)^{4}\bigr],
\qquad
\frac{\overline{\mathcal P}_{\mathrm{sheet}}}{\overline{\mathcal P}_{\mathrm{dip}}}
=\frac{(kR)^{4}}{100}+O\bigl[(kR)^{6}\bigr].
\label{eq:suppression}
\end{equation}
For finite $kR$, the ratio oscillates and reaches extrema at the stationary points of $j_2$,
\begin{equation}
x\simeq3.3421,\quad \tfrac94 j_2^{2}\simeq0.2118,
\qquad
x\simeq7.2899,\quad \tfrac94 j_2^{2}\simeq0.0438 .
\label{eq:extrema}
\end{equation}
The suppression follows from the dipole cancellation in Eq.~\eqref{eq:freeff}. The free source removes the bound source's point-dipole radiation exactly. The remaining field is a finite-size retardation term in the same electric-parity $\ell=1$ channel.
\\

The leading term also has a standard multipole interpretation. In the Cartesian expansion of Ref.~\cite{miroshnichenko}, the far field of a compact source contains a Cartesian electric dipole $\vp$ and a toroidal dipole, defined here by
\begin{equation}
\vT=\frac{1}{10}\int\Bigl[\bigl(\vx\cdot\vJ\bigr)\,\vx-2r^{2}\vJ\Bigr]\dd^{3}x,
\label{eq:toroidaldef}
\end{equation}
and these moments enter the far field through $\vp+(ik/c)\vT$. An equivalent convention absorbs the factor of $c$ into the moment by defining $\vT_{\mathrm{car}}=\vT/c$ and writing $\vp+ik\,\vT_{\mathrm{car}}$. For the neutralized minimal-sheet source, $\vp=\zero$ identically. The ball and coating contribute with opposite signs to Eq.~\eqref{eq:toroidaldef},
\begin{equation}
\vT=\frac{1}{10}\Bigl[-\frac{4\pi}{3}+\frac{8\pi}{3}\Bigr]R^{5}\,\frac{\dd P_{s}}{\dd t}\,\zhat
=\frac{2\pi}{15}\,R^{5}\,\frac{\dd P_{s}}{\dd t}\,\zhat,
\label{eq:toroidalvalue}
\end{equation}
where the first bracketed term comes from the volume current and the second from the sheet. A pure toroidal dipole radiates the point-dipole power in Eq.~\eqref{eq:dipref} with effective moment $(\omega/c^{2})\lvert\vT\rvert$, so
\begin{equation}
\overline{\mathcal P}_{T}
=\frac{\mu_{0}\omega^{4}}{12\pi c}
\Bigl(\frac{\omega}{c^{2}}\cdot\frac{2\pi}{15}\,\omega P_{0}R^{5}\Bigr)^{2}
=\frac{\pi\mu_{0}\omega^{8}R^{10}P_{0}^{2}}{675c^{5}},
\label{eq:toroidalpower}
\end{equation}
which agrees exactly with Eq.~\eqref{eq:Psmall}. Since the total charge density vanishes pointwise, the expansion contains no charge-weighted contribution. The matching coefficient identifies the leading field in Eq.~\eqref{eq:Esmall} as purely toroidal. The Cartesian electric dipole vanishes identically rather than through parameter tuning. At finite $kR$, the same electric-parity $\ell=1$ channel also contains higher current-multipole corrections, all resummed by the closed form $\tfrac32 j_{2}(kR)$.

\subsection{Nonradiating Frequencies}

The angular pattern has directional nodes at $\theta=0$ and $\pi$ for every frequency. A separate class of zeros occurs when the only exterior coefficient in Eq.~\eqref{eq:C1} vanishes, as specified by Eq.~\eqref{eq:zeros}. Bracketing the sign changes of $j_2$ and applying Brent's method with absolute tolerance $10^{-12}$ gives the first three positive roots,
\begin{equation}
 kR=5.763459,\quad 9.095011,\quad 12.322941.
\label{eq:j2zeros}
\end{equation}
At any of these roots, $C_1=0$. Equation~\eqref{eq:vanish} shows that all other coefficients already vanish, so the exterior field in Eq.~\eqref{eq:Bout} is identically zero. The source is exactly nonradiating, with $\vE=\vB=\mathbf 0$ throughout $r>R$. No exterior uniqueness theorem is required because the solution contains only one coefficient. Nonzero currents and fields remain inside. The Amp\`ere-Maxwell law forbids a nonzero current with $\vE=\vB=\mathbf 0$ everywhere, and Sec.~\ref{sec:exact} gives the interior fields explicitly. The interior field in Eq.~\eqref{eq:Bpair} contains the coefficient in Eq.~\eqref{eq:A1}, which has zero real part at a root but a nonzero imaginary part according to Eq.~\eqref{eq:A1root}. The interior electric field therefore has no quadrature component relative to the polarization, and the driving agent performs zero average work by Eq.~\eqref{eq:Pinresult}. In this setting, nonradiating means field-free outside the source rather than field-free everywhere. The first root gives the scale
\begin{equation}
\frac{R}{\lambda}=\frac{5.763459}{2\pi}\simeq0.917,
\qquad
2R\simeq1.83\,\lambda ,
\label{eq:size}
\end{equation}
which shows that the exact zeros are finite-size effects and do not belong to the dipole limit.
\\

The silence conditions can also be written in transcendental form. From Eq.~\eqref{eq:jdefs}, $j_{1}=\cos x\,(\tan x-x)/x^{2}$ and $j_{2}=\cos x\,[(3-x^{2})\tan x-3x]/x^{3}$. At their zeros, $\cos x$ and the displayed denominators remain nonzero. Therefore
\begin{equation}
j_{1}(x)=0\;\Longleftrightarrow\;\tan x=x,
\qquad
j_{2}(x)=0\;\Longleftrightarrow\;\tan x=\frac{3x}{3-x^{2}},
\label{eq:tanforms}
\end{equation}
\begin{equation}
j_{2}(x)=2j_{0}(x)\;\Longleftrightarrow\;\tan x=\frac{x}{1-x^{2}},
\qquad
x\simeq2.743707,\quad 6.116764,\ldots,
\label{eq:tanfree}
\end{equation}
with the final relation giving the silence sequence of the free subsystem alone from Eq.~\eqref{eq:freeclosed}. Its first roots are obtained by the same bracketing procedure. The bare sphere, the neutralized pair, and the free subsystem each become silent at intersections of a tangent function with a different rational function.
\\

The two root sequences of the complete radiating sources cannot coincide,
\begin{equation}
j_{1}(x)=j_{2}(x)=0
\;\Longrightarrow\;
j_{0}(x)=0
\;\Longrightarrow\;
x=n\pi,
\qquad
j_{1}(n\pi)=-\frac{\cos n\pi}{n\pi}\neq0 ,
\label{eq:nocommon}
\end{equation}
where the first implication follows from the recurrence $j_{2}=3j_{1}/x-j_{0}$. The contradiction proves that the bare-sphere and minimal-sheet cases in Table~\ref{tab:realizations} share no nonradiating frequency. These discrete zeros are a finite-size analogue of Schott's condition $2a=mcT$ for a rigidly orbiting charged shell~\cite{schott}. Schott's selected frequencies arise from the transit time of a moving geometry. Here the body is fixed, the polarization oscillates, and the boundary matching produces a coefficient proportional to $j_2(kR)$ with its own root sequence.

\subsection{Status of the Source Model}

The source is prescribed macroscopically rather than derived from a susceptibility, and the coating current is impressed. Feed lines and return paths are not included. The division into free and bound sources depends on the model, while only the total densities determine $\vE$ and $\vB$. A laboratory realization would introduce leads, finite coating thickness, and imperfect matching into $\vJ_{\mathrm{tot}}$. Each could add radiation. Equations~\eqref{eq:Erad} and \eqref{eq:power} describe the ideal prescribed current distribution.

\section{Concluding Remarks}
\label{sec:conclusion}

The neutralized sphere exposes a limitation of describing radiation through charge multipoles alone. A time-dependent charge density does not define the source until its current is also specified. In the static limit, the neutralized body has no $\vE$, no $\vB$, and no $\vH$ anywhere. It retains only a $\vD$ field equal to the impressed polarization inside. The two realizations are then indistinguishable because neither carries current. Once the polarization oscillates, the three complete sources have the exact form factors
\begin{equation}
F_{\mathrm{bare}}=\frac{3j_1(kR)}{kR},\qquad
F_{\mathrm{sheet}}=\frac{3}{2}\,j_2(kR),\qquad
F_{\mathrm{cancel}}=0,
\label{eq:hierarchy}
\end{equation}
for the bare sphere, the minimal tangential-sheet neutralization, and complete cancellation of charge and current, respectively. These factors describe the full exterior fields, not only their radiation-zone limits. Each radiating source is exactly equivalent outside the body to a point dipole whose effective moment is the bare moment $p_{b}$ multiplied by the corresponding form factor. The two neutralized sources share the same prescribed charge history but radiate differently. The minimal sheet source becomes nonradiating at the roots of $j_2$ because its exterior coefficient is proportional to $j_2(kR)$. Neutralization alone does not select those roots. The volume-current source is equally neutralized and remains nonradiating at every frequency. The energy balance is consistent with this picture. The average power supplied to the polarization and coating equals the radiated power, and it vanishes at roots of $j_2$ even though interior fields remain. Continuity determines the amount of current converging on a changing charge, but it does not determine the path that current follows. Radiation makes that freedom observable.

\FloatBarrier

\section*{Statements and Declarations}
\noindent\textbf{Funding}\quad No funding was received for this study.

\medskip
\noindent\textbf{Competing Interests}\quad The author declares no competing interests.

\medskip
\noindent\textbf{Ethics Approval and Consent to Participate}\quad Not applicable.

\medskip
\noindent\textbf{Consent for Publication}\quad Not applicable.

\medskip
\noindent\textbf{Data Availability}\quad No datasets were generated or analyzed. The figures are schematic or are direct numerical plots of the closed-form expressions given in the manuscript.

\medskip
\noindent\textbf{Materials Availability}\quad Not applicable.

\medskip
\noindent\textbf{Code Availability}\quad A Python script was used only to prepare the figures. The definitions, derivations, inputs, and conclusions are fully stated in the manuscript and do not depend on access to the script.

\medskip
\noindent\textbf{Author Contributions}\quad N.R. conceived the study, completed the analytical derivations, prepared the figures, and wrote the manuscript.
\newpage

\FloatBarrier

\end{document}